\documentclass[twocolumn]{elsart5p}
\usepackage{graphicx,amsmath,amsfonts,mathrsfs}
\usepackage{amssymb}
\usepackage{cite}

\def\lsim{\stackrel{\scriptstyle <}{\phantom{}_{\sim}}}

\def\rmd{{\rm d}}
\makeatletter
\let\old@ssect\@ssect % Store how ifacconf defines \@ssect
\makeatother
\usepackage[plainpages=false,pdfpagelabels=false]{hyperref}
\makeatletter
\def\@ssect#1#2#3#4#5#6{%
  \NR@gettitle{#6}% Insert key \nameref title grab
  \old@ssect{#1}{#2}{#3}{#4}{#5}{#6}% Restore ifacconf's \@ssect
}
\makeatother
\DeclareMathOperator{\rot}{rot}

\begin{document}

\begin{frontmatter}

\title{Hyperon global polarization in heavy-ion collisions at NICA energies.\\ Feed-down effects and the role of $\Sigma^0$ hyperons.}
%%%%%%%%%%%%%%%%%%%%%
\author[JINR,ITP]{V. Voronyuk}, %\and
\author[JINR,UMB]{E.E. Kolomeitsev}, \and
\author[JINR]{N.S. Tsegelnik}
%%%%%%%%%%%%%%%%%%%%%
\address[JINR]{Joint Institute for Nuclear Research (JINR), Dubna, Russia}
\address[ITP]{Bogolyubov Institute for Theoretical Physics, Kiev, Ukraine}
\address[UMB]{Matej Bel University,  Banska Bystrica, Slovakia}
\begin{abstract}
Global polarization of hyperons induced by the local vorticity of the medium created in heavy-ion collisions at energies 2.3\,GeV$\le\sqrt{s_{NN}}\le$11.5\,GeV is calculated in the parton-hadron-string dynamic (PHSD) model. The separation of spectator nucleons and the fluidization of the generated particle distributions are performed. The polarization of all anti-hyperon species is found significantly larger than that of hyperons. The $\overline{\Xi}$ hyperons are found to be polarized as strong as $\overline{\Lambda}$s but $\Xi$ hyperons have weaker polarization compared to $\Lambda$s. The $\Omega$ and $\overline{\Omega}$ polarizations show the strongest dependence on the collision energy. Despite the strong polarization of the produced $\Lambda$s and $\overline{\Lambda}$s induced by the vortical flows in the medium, the observed polarization signal is significantly depleted because of the feed down from weak and electromagnetic decays of heavier hyperons. Particularly strong suppression is found to be due to electromagnetic decays of $\Sigma^0$ hyperons, which multiplicities obtained in the transport are poorly constrained both from the microscopic input of the $\Sigma^0$ production reactions and from the experimental data. The final $\Lambda(\overline{\Lambda})$ polarization signal strongly depends on the $\Sigma^0$ multiplicity generated in the model. With all these effects we can reproduce the measured global $\Lambda$ polarization in collisions at $\sqrt{s_{NN}}=7.7$ and $11.5$\,GeV and the global $\overline{\Lambda}$ polarization at 11.5\,GeV. For energies $\lsim 3$\,GeV, the calculated $\Lambda$ polarization is smaller than the observed one.   The polarization of $\Xi(\overline{\Xi})$ hyperons is calculated. The signal of $\Xi(\overline{\Xi})$ polarization is argued to be insensitive to feed-down effects and be, therefore, a more direct probe of the degree of the vorticity in the system.
\end{abstract}


\end{frontmatter}

%%%%%%%%%%%%%%%%%%%%%%%%%%%%%%%%%%%%%%%%
\section{Introduction}

The non-vanishing global spin polarization of hyperons observed by the STAR~\cite{Abelev:2007, Adamczyk:2017, PhysRevC.104.L061901, Adam:2021} and HADES~\cite{Yassine:2022} collaborations raises new theoretical questions. Why does the global spin polarization of hyperons persists in a heavy-ion collision where the reaction plane of an elementary hard nucleon-nucleon interaction is randomly oriented? If the polarization is related to the total angular momentum accumulated in the medium, why then the degree of polarization increases with a decrease of collision energy, i.e., for a smaller initial angular momentum of colliding nuclei? Finally, why the polarization of anti-hyperons is systematically larger than those of hyperons, whereas, in high-energy $p+p$ and $p+A$ collisions, anti-hyperons get zero polarization~\cite{DeGrand-81}?

Various mechanisms were proposed to explain the observed hyperon polarization. Within the statistical approach~\cite{Becattini:2013,Becattini:2015ska} the polarization was linked with the thermal vorticity distributed in the fireball medium. At the moment, this is the most studied source of the particle polarization in heavy-ion collisions (HICs) in which the initial angular momentum of colliding nuclei is partially transferred to the rotation of the medium~\cite{PhysRevC77.024906}. This mechanism is the relativistic generalization of the Barnett effect~\cite{Barnett}, which manifests itself in the magnetization of an uncharged body through rotation. The statistical approach is most naturally realized in the hydrodynamic calculations~\cite{Becattini:2015ska,Karpenko:2017,Xie:2017,Ivanov-Soldatov-2017,Ivanov-Toneev-Soldatov-2019,Ivanov:2020,Ivanov:2021}.

Other contributions to the polarization due to the thermal shear and the analog of the spin-Hall effect were proposed in Refs.~\cite{Becattini:2021,Liu:2021,Becattinit:2021:2} and Ref.~\cite{Liu:2021:2}, respectively. There are some ambiguities, however, in the inclusion of these effects in hydrodynamic codes~\cite{Ivanov-zhetp}. The statistical mechanism of the spin polarization generation was applied also in the transport codes: AMPT in Refs.~\cite{Sun:2017,Li:2017,Han-Xu-18,Wei:2019,Li-Xia-Huang-Huang-22,Xu-Lin-Huang-Huang-22},
UrQMD~\cite{VITIUK2020135298,Deng-Huang-Ma-22}, QGSM~\cite{BGST-Hseparation,BGST-Vsheet}, and PHSD~\cite{PhysRevC.97.064902}. Although fluidization procedures for conversion of the particle coordinate and momentum distributions, generated in the code, into velocity and temperature fields may differ significantly from one work to another, most of these calculations can successfully describe $\Lambda$ polarization at energies above $\sqrt{s_{NN}}=7$\,GeV. The numerical simulations confirm that the vorticity field may be quite developed in HICs and have intricate space-time structures, such as vortex sheets (femtocyclones)~\cite{BGST-Hseparation,BGST-Vsheet} or elliptic vortex rings~\cite{Ivanov-rings,helicity}. Nevertheless, the large degree of spin polarization at low collision energies observed by HADES remains quite challenging. Also, the $\overline{\Lambda}$--$\Lambda$ splitting can not be reproduced in almost all calculations unless a specific mechanism distinguishing particles from anti-particles is incorporated. To such mechanisms one counts: the magnetic fields created in the collision~\cite{Voskre-mag,SIT-2009}, which influence on the polarization was considered in~\cite{Becattini:2017,Han-Xu-18,Buzzegoli-mag-2022}; the chiral or axial anomaly~\cite{Rogachevsky-ST-2010,Gao:2012} included in the transport code calculations in~\cite{BGST-Hseparation,BGST-Vsheet,Sun:2017,Baznat:2018}; the influence of the `magnetic parts' of mean vector fields induced by vorticity~\cite{Csernai:2019} and included in hydrodynamic calculations in~\cite{Xie-Chen-Csernai-2021,Ivanov:2022}.
The $\Lambda$--$\overline{\Lambda}$ splitting can be also obtained in the geometrical approach distinguishing central and peripheral collisions,
the so-called core-corona mechanism~\cite{Ladygin:2010,Ayala-PLB810,Ayala-Particles23}. Without these additional effects, the splitting was reproduced only in Ref.~\cite{VITIUK2020135298} where it was attributed to the different space-time distributions of $\Lambda$ and $\overline{\Lambda}$ and by different freeze-out conditions of hyperons and anti-hyperons. Indeed, the highest vorticity modulus is argued in Refs.~\cite{Deng-Huang-Ma-Zhang-20,PhysRevC.97.064902} to correspond to the initial stage of the system evolution, when the medium is hot and occupies a small spatial volume. The vorticity gradually decays with the fireball expansion. Hence depending on the typical freeze-out time of particles, they feel different vorticity fields and different spin polarization signals may be formed.
Applying the PHSD transport code, the interconnection between $\Lambda$-$\overline{\Lambda}$ freeze-out distribution and spin polarization was
investigated in Ref.~\cite{TKV-particles-23}. It was shown that the global polarizations for both $\Lambda$s and $\overline{\Lambda}$s measured by the STAR collaboration can be quantitatively reproduced at energy $\sqrt{s_{NN}} = 11.5$\,GeV. For collisions at $\sqrt{s_{NN}} = 7.7$\,GeV, the $\Lambda$ polarization was also reproduced, but the $\overline{\Lambda}$ polarization was significantly underpredicted.

In this work, we apply the approach described in Refs.~\cite{helicity,TKV-particles-23} to systematic analysis of the collision energy dependence of the $\Lambda$ and $\overline{\Lambda}$ polarization. Also, the polarization of other hyperon species will be investigated. Particular emphasis should be put on the feed-down effects, which are important for the formation of the final polarization signal.

The paper is organized as follows. In Sect.~\ref{sec:hyperon-product} we analyze the production rate and the freeze-out time for different hyperon species. The excitation functions for various hyperons and anti-hyperons are compared with experimental data.
In Sec.~\ref{sec:vorticity} the global spin polarization for various hyperon species is calculated. The feed-down effects due to weak and electromagnetic decays are considered in Sect.~\ref{sec:feed-down}. Conclusions are formulated in Sect.~\ref{sec:conclusion}.

%\clearpage
\section{Hyperon production}\label{sec:hyperon-product}

\begin{figure}
\centering
\includegraphics[width=8cm]{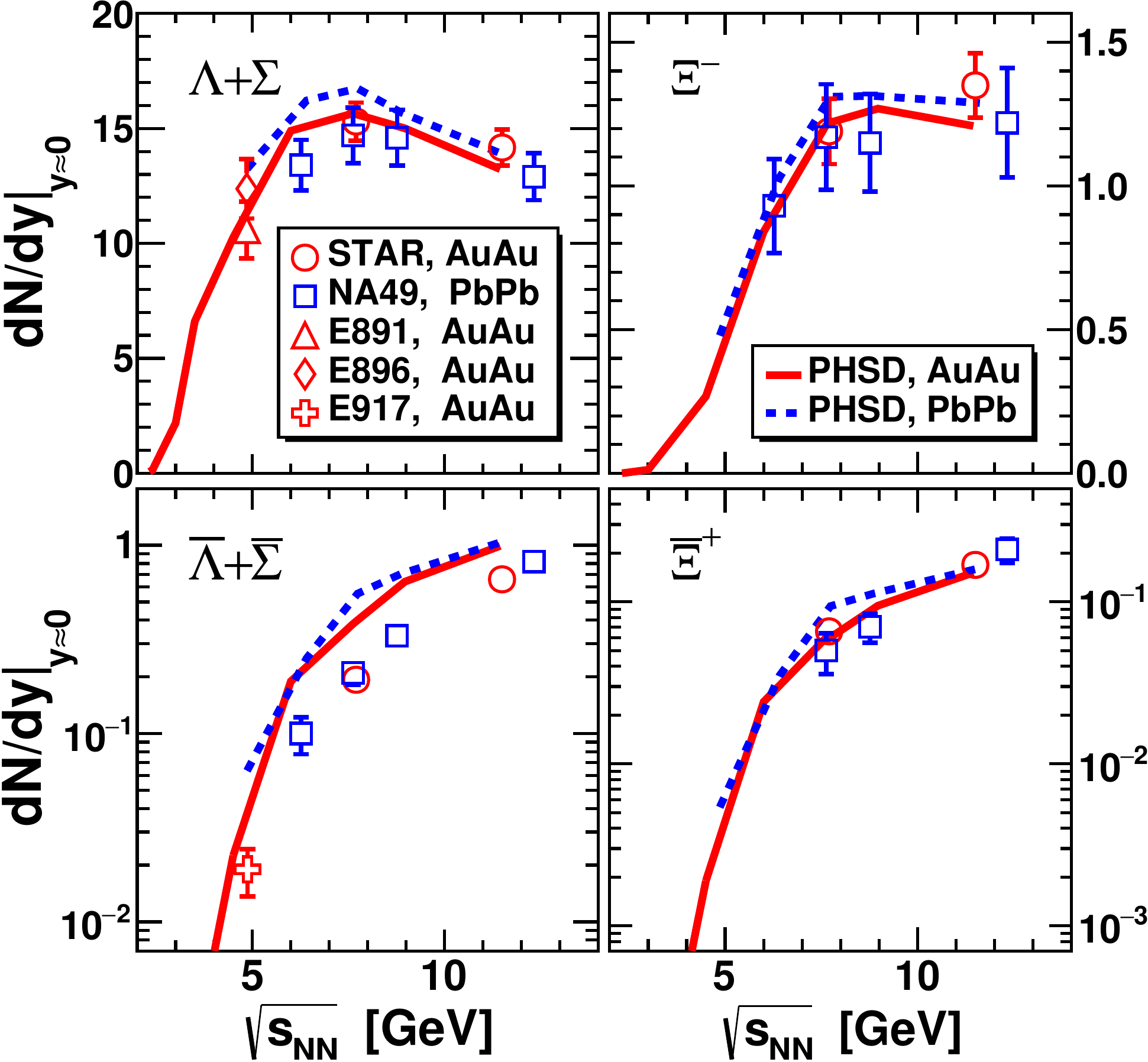}
\caption{The collision energy dependence of the $\Lambda+\Sigma$ and $\overline{\Lambda}+\overline{\Sigma}$ (left), $\Xi^-$ and $\overline{\Xi}^+$ (right) yields at mid-rapidity in central Au+Au and Pb+Pb collisions at $\sqrt{s_{NN}}=2-14$\,GeV.
Calculations are performed for the centrality class 0--7\%. The $\Lambda$ and $\overline{\Lambda}$ results of AGS~\cite{PhysRevB.386.034909, PhysRevLett.87.242301, PhysRevLett.88.062301} are inclusive, and those of NA49~\cite{PhysRevC.78.034918} and STAR~\cite{PhysRevC.102.034909} are corrected for the weak decay feed-down.}
	\label{fig:dNdY-Y0-LambdaXi}
\end{figure}

The PHSD transport model~\cite{PHSD-I,PHSD-II,PHSD-III} proved to be successful in the description of strangeness production in HICs. In the current version, the chiral symmetry restoration effects~\cite{Cassing:2015owa} are taken into account which increases the probability of the strangeness production in initial hard processes. This allows for a good description of the strange particle yields at AGS and SPS energies~\cite{PHSD-IV}. In Fig.~\ref{fig:dNdY-Y0-LambdaXi} we show the hyperon ($\Lambda+\Sigma^0$ and $\Xi$) and anti-hyperon
($\overline{\Lambda}+\overline{\Sigma}^0$ and $\overline{\Xi}$) multiplicities in Au+Au and Pb+Pb collisions at mid-rapidity as functions of the collision energy in comparison with experimental data~\cite{PhysRevB.386.034909, PhysRevLett.87.242301, PhysRevLett.88.062301,PhysRevC.78.034918,PhysRevC.102.034909}. We observe good overall agreement between theory and experiment except for the multiplicity of $\overline{\Lambda}+\overline{\Sigma}$ being overestimated.

The only data on the $\Omega$ ant $\overline{\Omega}$ multiplicities at low energies were reported by the NA49 collaboration~\cite{NA49-Alt-Omega}  for central Pb+Pb collisions at energy 40$A$\,GeV (i.e., $\sqrt{s_{NN}}=8.77$\,GeV)
\begin{align}
M_{\Omega + \overline{\Omega}}^{\rm (exp)} = 0.14\pm 0.05\,.
\label{NOmega-exp}
\end{align}
Our calculations for centrality class 0--7\% give
\begin{align}
M_{\Omega} = 0.123 \,,\quad  M_{\overline{\Omega}} = 0.018\, .
\end{align}
The sum of these multiplicities agrees nicely with the experimental value (\ref{NOmega-exp}). The transverse momentum distributions for $\Lambda+\Sigma^0$ and $\overline{\Lambda}+\overline{\Sigma}^0$ are presented in Ref.~\cite{TKV-particles-23}  for Au+Au collisions at $\sqrt{s_{NN}}=7.7$\,GeV
and various centralities. The agreement with the data obtained by the STAR collaboration is quite satisfactory.

\begin{figure}
\centering
\includegraphics[width=6cm]{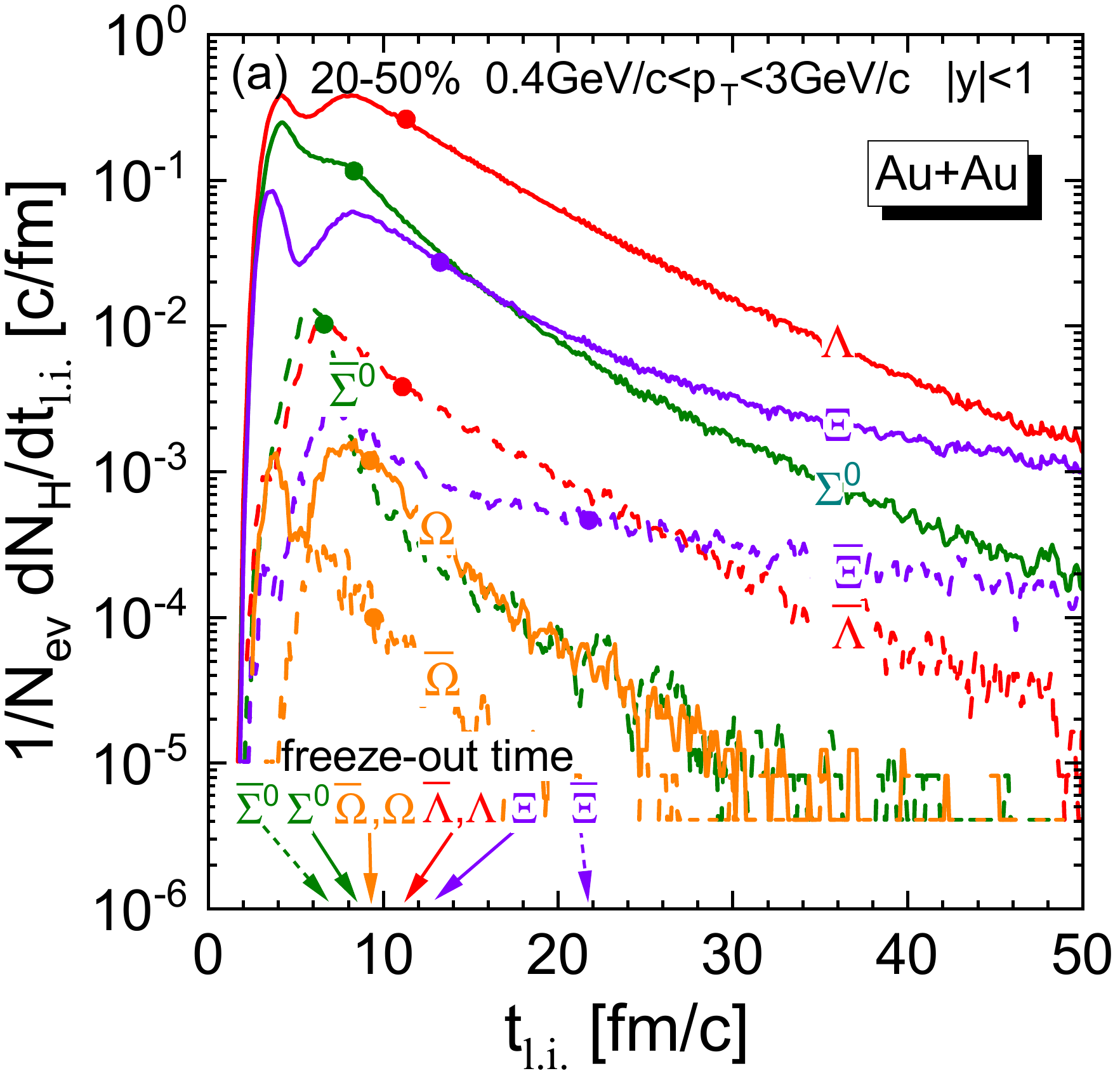}\\
\includegraphics[width=6cm]{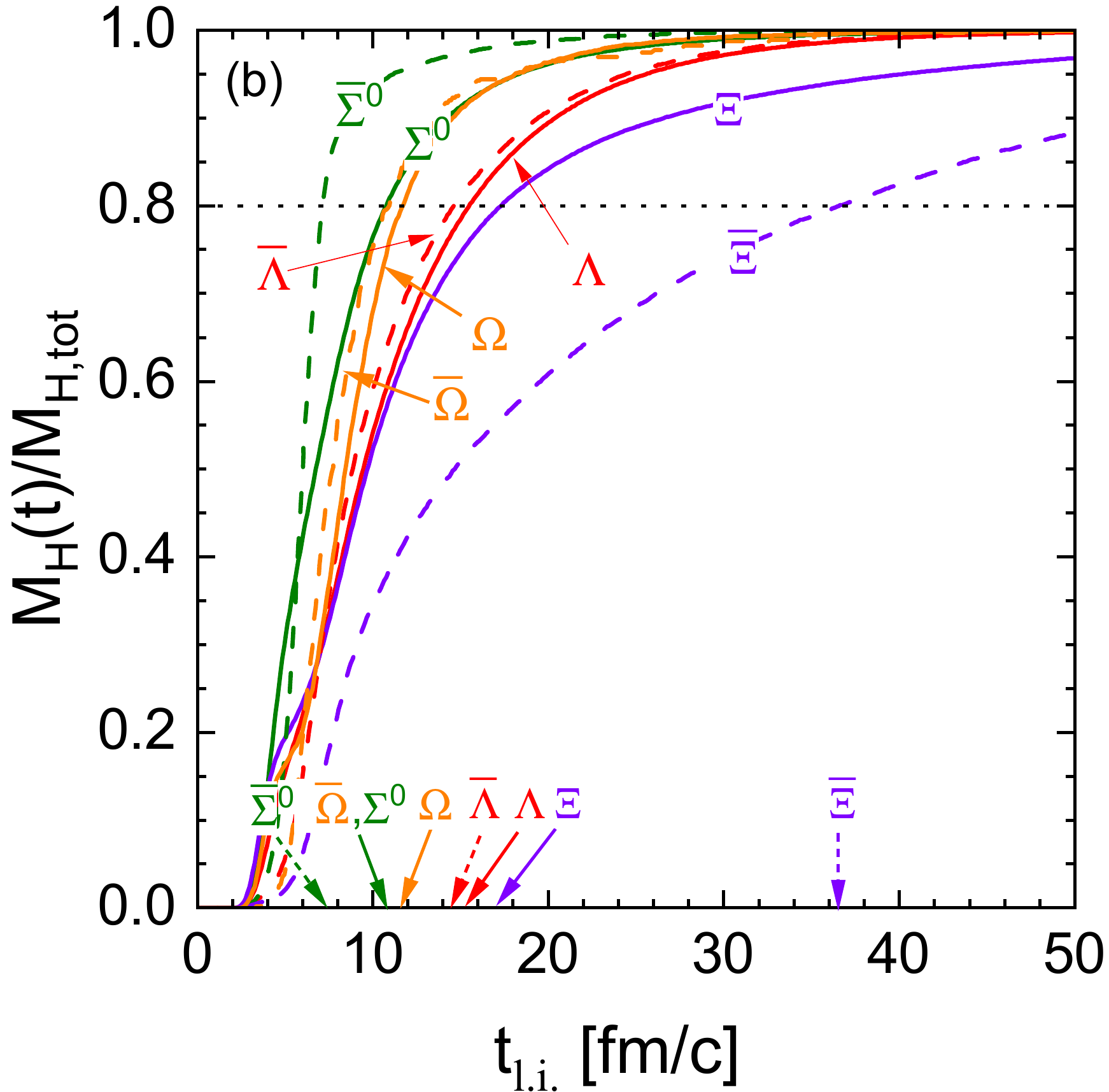}
\caption{
Panel~(a): Rates of the hyperons production $N_{\rm ev}^{-1}\rmd N_H/\rmd t_{\rm l.i.}$ normalized by the total number of events $N_{\rm ev}$ as functions of the time of the last interaction, $t_{\rm l.i.}$.  Arrows mark the mean freeze-out times calculated according to Eq.~\eqref{eq:freeze-out}. Panel~(b): The multiplicity of released hyperons $M_{\rm H}$ as functions of $t_{\rm l.i.}$ normalized by the total multiplicity $M_{\rm H,tot}$ of produced hyperons $H$, see Eg.~(\ref{MH-def}). Arrows mark the times at which $80\%$ of hyperons are released.
Calculations on both panels are performed for Au+Au collisions at energy $\sqrt{s_{NN}}=7.7$\,GeV for mid-rapidity $|y|<1$ and the transverse momentum range $0.4<p_{T}<3\,$GeV/$c$.}
\label{fig:NTimeLi}
\end{figure}

In the present version of the code at each time step we store the time marker for each 'newly-created' particle, i.e., the particle formed either in the string breaking or hadronic rescatterings, or a strong resonance decay. After the completion of a code run, when momentum distributions of particles stop changing (for the studied energy range it takes about $t_{\rm fin}\sim 60-100$\,fm/$c$ in total), we can look at hyperons that survive all strong interactions and can, eventually, be caught by a detector, and obtain the distribution of the time of the last interaction, $t_{\rm l.i.}$ (hereinafter, TLI). Such distributions for all hyperon species averaged over $N_{\rm ev}$ code runs, $\frac{1}{N_{\rm ev}}\frac{\rmd N_H}{\rmd t_{\rm l.i.}}$ are shown in Fig.~\ref{fig:NTimeLi}a for Au+Au collisions at $\sqrt{s_{NN}}=7.7$\,GeV.
We apply cuts in rapidity and transverse momentum and select centralities according to the experimental works on measuring the polarization of $\Lambda$-hyperons~\cite{Abelev:2007,Adamczyk:2017,PhysRevC.104.L061901}. The zero-time moment in Fig.~\ref{fig:NTimeLi} corresponds to the initialization of the nuclei. The touching time of gold nuclei is about $t \approx 2.2$~fm/$c$, the time of the maximal overlap of nuclei is $t \approx 4.9$\,fm/$c$ at $\sqrt{s_{NN}}=7.7$~GeV. We see the qualitative difference between the hyperon and anti-hyperon distributions: there are two
maxima in the distribution for hyperons and only one for anti-hyperons. The first maximum for hyperons at times $t^{(1)}_H\simeq 4\mbox{--}5\,$fm/$c$ corresponds to the hyperon production in initial hard collisions whereas the second maximum at $t^{(2)}_H\simeq 8\,$fm/$c$ can be associated with the hyperon production in hadron collisions, resonance decays, and parton hadronization. The times of the first and second maxima differ for various hyperons by $\sim 0.5$\,fm/$c$ and have the following hierarchy $t_\Xi^{(1)} < t_\Omega^{(1)}< t_{\Sigma^0}^{(1)} \sim t_\Lambda^{(1)}$ and $t_{\Sigma^0}^{(2)} \sim t_{\Lambda}^{(2)} < t_\Xi^{(2)}\sim t_\Omega^{(2)}$.
On the contrary, the TLI distributions have only one maximum for anti-hyperons, which lies typically between two maxima for hyperons, and
$t_{\overline{\Sigma}^0} \simeq 5.7\,{\rm fm}/c< t_{\overline{\Omega}} < t_{\overline{\Xi}} < t_{\overline{\Lambda}}\simeq 6.9$\,fm/$c$.

In Ref.~\cite{VITIUK2020135298} the authors used the mean freeze-out time for comparison of hyperon and anti-hyperons production times. We define the corresponding mean time  when the particles are released from the fireball as
\begin{equation}\label{eq:freeze-out}
t_H^{\rm (f.o.)}=\int_{0}^{t_{\rm fin}} t_{\rm l.i.} \frac{\rmd N_{\rm H}}{\rmd t_{\rm l.i.}} \rmd t_{\rm l.i.}\Big/
\int_{0}^{t_{\rm fin}} \frac{\rmd N_{\rm H}}{\rmd t_{\rm l.i.}} \rmd t_{\rm l.i.}\,.
\end{equation}
This quantity is sensitive to the decrease rate of the $\frac{\rmd N_{\rm H}}{\rmd t_{\rm l.i.}} $ distribution for  increasing time marker $t_{\rm l.i}$. The slower the falloff of the distribution is, the larger will be $t_H^{\rm (f.o.)}$.
The mean freeze-out times are denoted by solid circles in Fig.~\ref{fig:NTimeLi}a. We see that $t_H^{\rm (f.o.)}$ and $t_{\overline{H}}^{\rm ( f.o.)}$ are very close for $\Lambda$ and $\Omega$ species, $t^{\rm (f.o.)}_\Lambda\simeq t^{\rm (f.o.)}_{\overline\Lambda}\simeq 11$\,fm/$c$ and
$t^{\rm (f.o.)}_\Omega\simeq t^{\rm (f.o.)}_{\overline\Omega}\simeq 9$\,fm/$c$. For $\Sigma^0$ and $\overline{\Sigma}^0$ we find
$t^{\rm (f.o.)}_{\Sigma^0}\simeq 8$\,fm/$c$ and $t^{\rm (f.o.)}_{\overline{\Sigma}^0}\simeq 6$\,fm/$c$.
The largest difference is, however, found  for $\Xi$ and $\overline{\Xi}$ hyperons: $t^{\rm (f.o.)}_{\Xi}\simeq 13$\,fm/$c$ and
$t^{\rm (f.o.)}_{\overline{\Xi}}\simeq 22$\,fm/$c$.
In comparison with calculations in~\cite{VITIUK2020135298} done within the  UrQMD model, we find a smaller difference in the mean freeze-out times for $\Lambda(\overline{\Lambda})$ hyperons $t^{\rm (f.o.)}_\Lambda - t^{\rm (f.o.)}_{\overline\Lambda}\simeq 0.27$\,fm/$c$ against $\sim 1$\,fm/$c$ in~\cite{VITIUK2020135298}, whereas the mean time itself is twice as short $t^{\rm (f.o.)}_{\Lambda(\overline{\Lambda})}\simeq 11$\,fm/$c$. In general the interacting phase of the nucleus-nucleus collision last in the UrQMD model longer than in the PHSD model.

Integrating the distribution shown in Fig.~\ref{fig:NTimeLi}a over the time, we obtain the hyperon multiplicities (the averaged number of hyperons in one collision event) accumulated to the time $t_{\rm l.i.}$
\begin{align}
M_H(t_{\rm l.i.})=\int_0^{\rm t_{l.i.}}\rmd t_{\rm l.i.}\frac{1}{N_{\rm ev}}\frac{\rmd N_H}{\rmd t_{\rm l.i.}}.
\label{MH-def}
\end{align}
The evolution of the hyperon multiplicity normalized by the final total multiplicity $M_{H,{\rm tot}}=M_H(t_{\rm fin})$ is presented in Fig.~\ref{fig:NTimeLi}b as functions of TLI. This plot allows to quantify the relative particle yields at various moments of time. Particularly, we see that the number of hyperons created at a later stage of collision is larger than the number of hyperons from hard processes. For orientation purposes we draw in Fig.~\ref{fig:NTimeLi}b the line corresponding to the release of $80\%$ of particles (dotted line).  The yields of $\Lambda(\Omega)$s and $\overline{\Lambda}(\overline{\Omega})$s grow at a similar pace, whereas the accumulation times for $\Sigma^0$ and $\overline{\Sigma}^0$ and, especially, for  $\Xi$ and $\overline{\Xi}$ are essentially different, so,
$t^{(80\%)}_\Xi\simeq 17.5$\,fm/$c$ while $t^{(80\%)}_{\overline{\Xi}}\simeq 37$\,fm/$c$.

Although the time-dependence of the $\Lambda$ and $\overline{\Lambda}$ yields is very similar, the thermodynamics conditions of the medium, wherein the hyperons have their last interactions, proved to be very different~\cite{TKV-particles-23}. In Ref.~\cite{TKV-particles-23} we argued that this difference leads to a difference in global spin polarizations for hyperons and anti-hyperons. The same conclusion was drawn also in Ref.~\cite{VITIUK2020135298}.

\begin{figure}
\centering
\includegraphics[width=9cm]{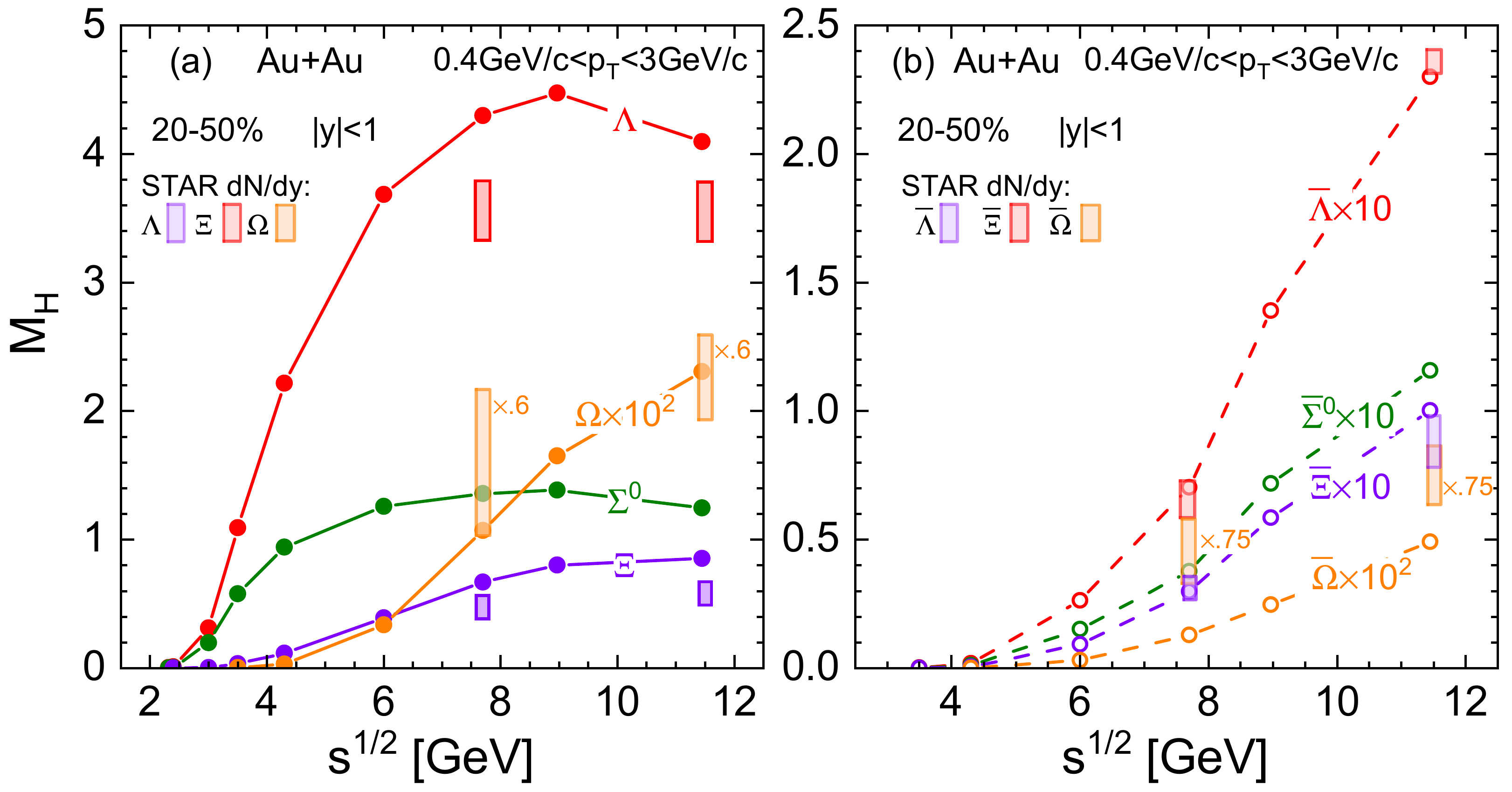}
    \caption{Hyperon (panel a) and anti-hyperon (panel b) multiplicities as functions of the collision energy for Au+Au collisions at centrality class 20--50\%, rapidities $|y|<1$, and the transverse momentum range $0.4\,{\rm GeV}/c< 3\,{\rm GeV}/c$. The results for $\Xi$ hyperons stand for the sum $\Xi^0 +\Xi^-$, the same is assumed also for the $\overline{\Xi}$ hyperons. The colored bars indicate the midrapidity multiplicities $\rmd M_H/\rmd y$ measured by the STAR collaboration~\cite{PhysRevC.102.034909} for $|y|<0.5$. See text for details.
\label{fig:Mhyp}}
\end{figure}

Concluding this section we show in Fig.~\ref{fig:Mhyp} the total multiplicities of strange baryons and anti-baryons produced in Au+Au collisions for centrality range and momentum cuts applied in the STAR polarization measurement~\cite{Adamczyk:2017}. Unfortunately, there are no experimental data on multiplicities obtained for the same centralities and momentum cuts. Just as a qualitative comparison we use the data on multiplicities at midrapidities reported in~\cite{PhysRevC.102.034909}. Selecting centrality class 20--60\% with the averaged number of participant-nucleons $\langle N_{\rm part}\rangle\simeq 110$ we can take the data for $\Lambda(\overline{\Lambda})$ and $\Xi^-(\overline{\Xi}^+)$ shown in Fig.~17 there. For $\Omega$ hyperons there are the data only for $\langle N_{\rm part}\rangle=153$, so we scaled the data down by factor 0.6 which follows approximately from the $\langle N_{\rm part}\rangle$ dependence of $\Lambda$, $\Xi$, and $\phi$ multiplicities. The experimental data for $\overline{\Omega}$ we multiplied by factor 0.75 to account for the differences between $\langle N_{\rm part}\rangle=123$ and 110.
We see that our calculations are compatible with the available experimental data for $\Lambda(\overline{\Lambda})$ and $\Xi^-(\overline{\Xi}^+)$, however, it seems that the $\Omega(\overline{\Omega})$ data remain underestimated. The latter could be also a signal that the $\Omega(\overline{\Omega})$ data decrease with the centrality increase faster than we estimated.

\section{Vorticity and global hyperon polarization}\label{sec:vorticity}

We will apply the statistical approach~\cite{Becattini:2013,Becattini:2015ska} to evaluate the average polarization of hyperons.
Within this approach, a local polarization is induced by a local thermal vorticity
\begin{align}
\label{eq:becattini:thermal-vorticity}
\varpi_{\mu\nu} = \frac{1}{2} (\partial_{\nu} \beta_{\mu} - \partial_{\mu} \beta_{\nu}), \quad \beta_{\nu} = \frac{u_{\nu}}{T}
\end{align}
where $u_\nu$ is the hydrodynamic four-velocity of the fluid element, and $T$ is the fluid temperature.
In leading order in $\varpi_{\mu\nu}$, the spin vector $S^{\mu}$ of a fermion with mass $m$ and spin $s=\frac12,\frac32,\dots$ is defined as
\begin{align}\label{eq:becattini:S-def}
S^{\mu}(x,p)=-\frac{s\, (s+1)}{6\, m}(1- n(x,p))\varepsilon^{\mu\nu\lambda\delta}\varpi_{\nu\lambda}p_\delta,
\end{align}
where  $p^{\mu}$ is 4-momentum of the fermion, and $ n(x,p) $ stands for the fermion distribution function.

Compared to our previous paper~\cite{PhysRevC.97.064902} with isochronous freeze-out, the method of the fluidization of the particle distributions generated in the code is significantly revised. First, we separate the spectator nucleons by the condition $||y| - y_{\rm beam}| \leq \Delta y$, where $y_{\rm beam}$ is the beam rapidity and $\Delta y=0.27$ is the Fermi-motion rapidity of nucleons in a nucleus. The spectator nucleons as the ``cold'' (not excited) matter do not form a fluid and, therefore, do not contribute to the thermal vorticity. The hydrodynamic velocity is calculated in the Landau frame, and the temperature is determined by the hadron resonance gas model~\cite{SDM09}.
Applicability of the hydrodynamic description is controlled by the condition for the local energy density $\epsilon> 0.05\,{\rm GeV/fm^3}$. For smaller energy densities, we assume that the medium does not exist, and, therefore, the temperature cannot be determined, and the thermal vorticity is put zero. The details of the thermodynamic characteristics of the medium, the velocity and vorticity fields created in Au+Au collisions at energies $\sqrt{s_{NN}}=4.5\mbox{--}11.5$\,GeV can be found in Refs.~\cite{helicity,TKV-particles-23}.

\begin{figure}
\centering
\includegraphics[width=6cm]{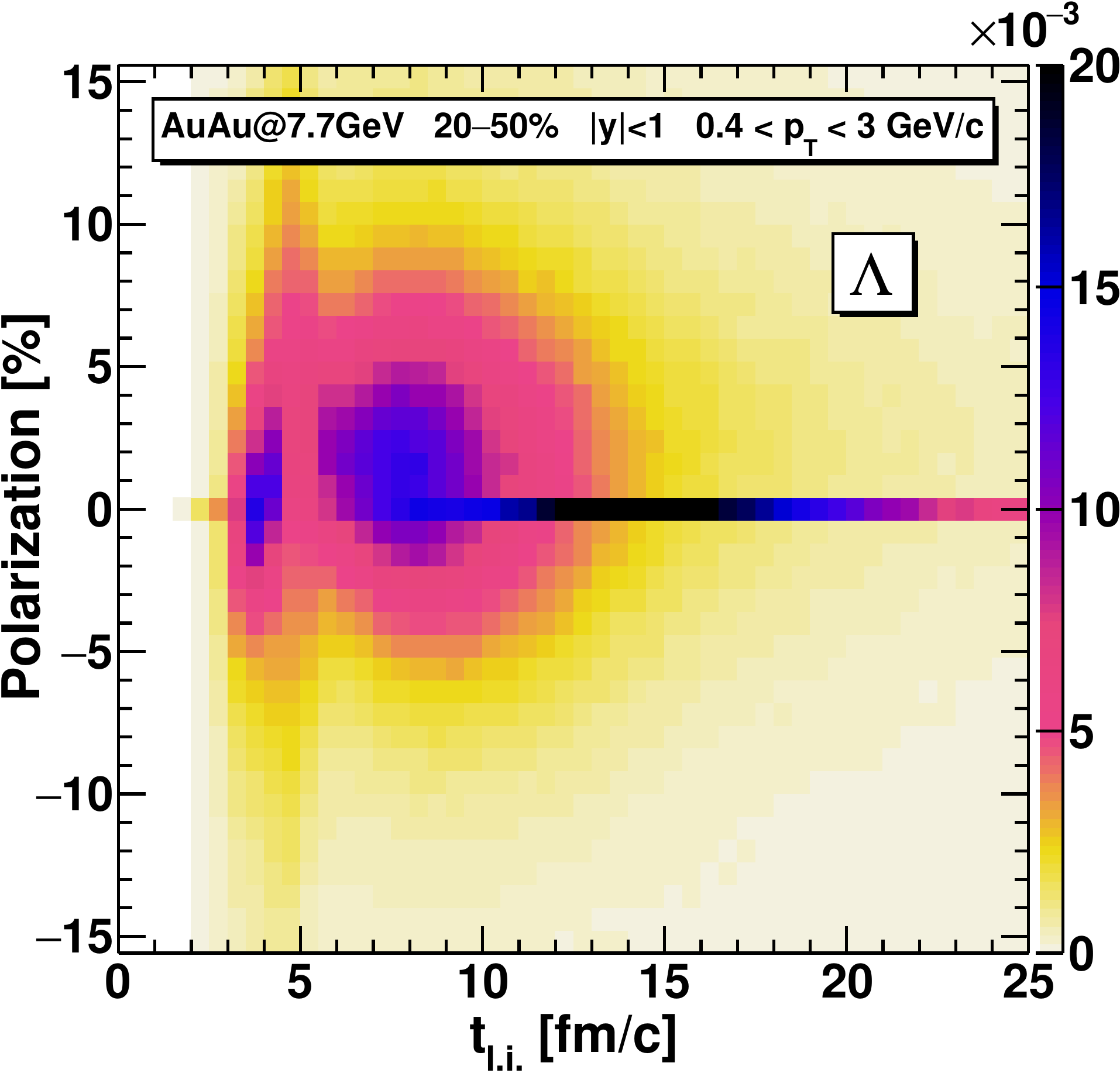}\\
\includegraphics[width=6cm]{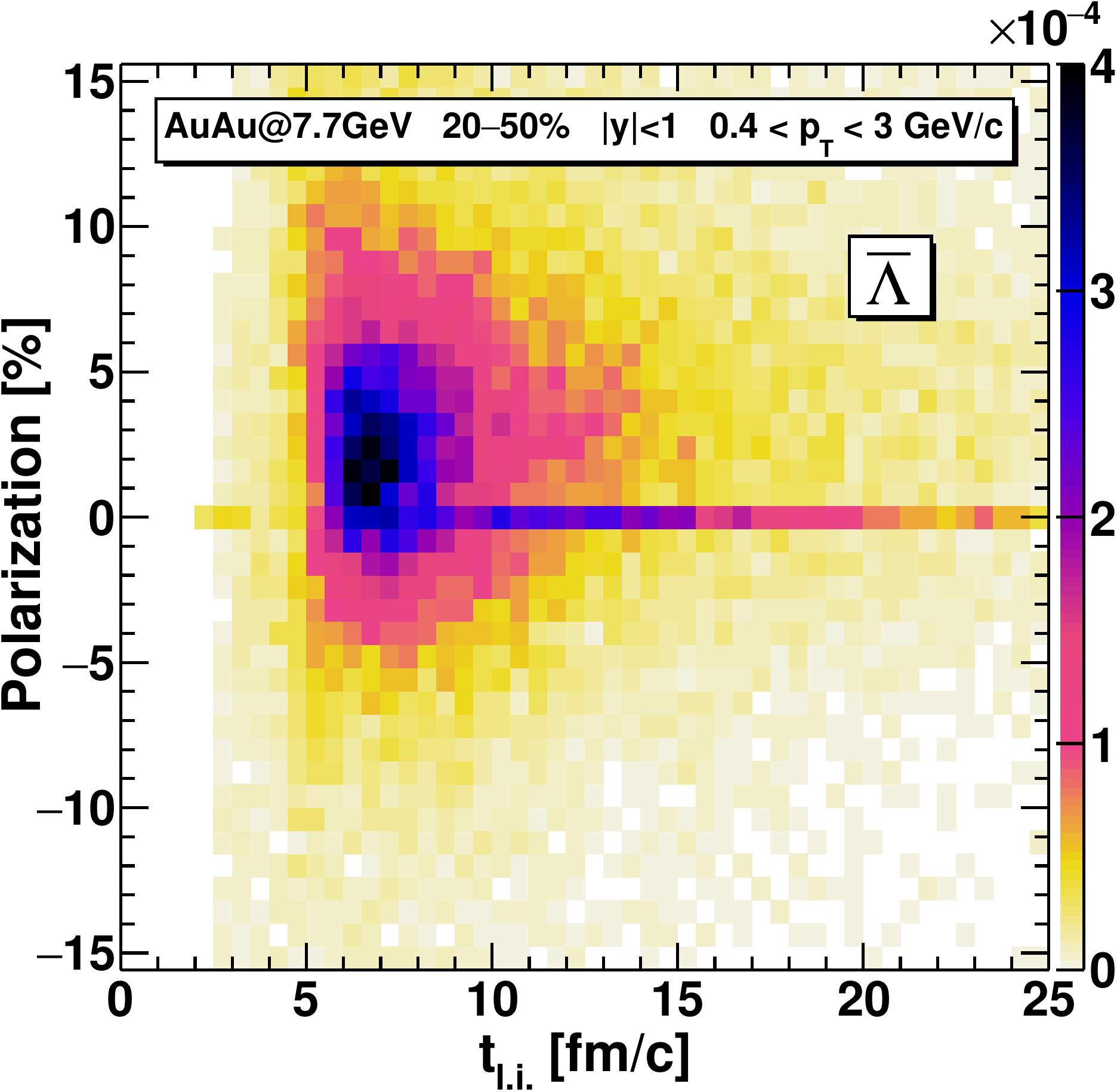}
\caption{
Distributions of the spin polarization of $\Lambda$ hyperons (upper panel) and $\overline{\Lambda}$ hyperons (lower panel), $w_H(t_{\rm l.i.}, P)$, see Eq.~(\ref{dW-dist}) as a function of the last interaction time, $t_{\rm l.i}$. Calculations are done for Au+Au collisions at energy $\sqrt{s_{NN}}=7.7$\,GeV. The centrality range and the transverse momentum and rapidity cuts are the same as in the STAR experiment~\cite{Adamczyk:2017}. The time of the maximum overlap of colliding nuclei is $t\approx5$\,fm/$c$.}
\label{fig:pol:time}
\end{figure}

Having determined the local thermal vorticity in the fluid cell with the center at coordinate $x$, we calculate the spin vector of a fermion with momentum $p$ using Eq.~(\ref{eq:becattini:S-def}). The contribution of the fermion to the averaged global polarization is given by $\vec{P^*}=\vec{S}^*/s$ where $\vec{S}^{*}$ is the spin vector (\ref{eq:becattini:S-def}) boosted to the fermion rest frame
\begin{align}\label{eq:becattini:S-boosted-vec}
\vec{S}^* = \vec{S} - (\vec{S} \vec{p}) \frac{\vec{p}}{E(E+m)}.
\end{align}
Note that in the rest frame $S_0^*=0$.
Taking into account that the Boltzmann limit of the particle momentum distributions is a good approximation in nucleus collisions at relativistic energies, i.e., $1- n(x,p) \approx 1$ we can write
\begin{align}
\vec{S}^* &\approx \frac{s(s+1)}{6m }
\Big( E \vec{\varpi}
+ [\vec{p}\times \vec{\varpi}_0]
-\frac{(\vec{p}\cdot \vec{\varpi})\vec{p}}{(E+M)}
\Big)\,,
\nonumber\\
(\vec{\varpi}_0)^i &=2 \varpi_{0i}=\partial_t\vec{\beta}+\vec{\nabla}\beta_0\,,\,
(\vec{\varpi})^i= \epsilon^{ijk}\varpi^{jk}=\rot \vec{\beta}\,.
\label{S*}
\end{align}
At each time step when a (anti-)hyperon is created or rescattered the spin polarization of the particle is calculated according to~Eqs.~\eqref{eq:becattini:S-def} or (\ref{S*}) if the fluid medium 'exists', i.e., the local energy density $\epsilon> 0.05\,{\rm GeV/fm^3}$.
In the absence of a medium, any hadronic collision (elastic or inelastic) nullifies the polarization.
If a hyperon is created in the strong decay,
\begin{align}
\label{eq:feedown:strong}
\Sigma^{*} \rightarrow \Lambda + \pi,\quad
\Xi^{*} \rightarrow \Xi + \pi,
\end{align}
the daughter hyperon inherits the direction of polarization of the parent hyperon resonances, but the magnitude is scaled by the factor
$C_{\Lambda\, \Sigma^*}=C_{\Xi\, \Xi^*} =1/3$ according to~\cite{Becattini:2017}.  Finally, after the nucleus collision ends and all strong interactions cease, the hyperons that have gone to infinity carry information about the medium and polarization at TLI.

\begin{figure}
\centering
\includegraphics[width=6cm]{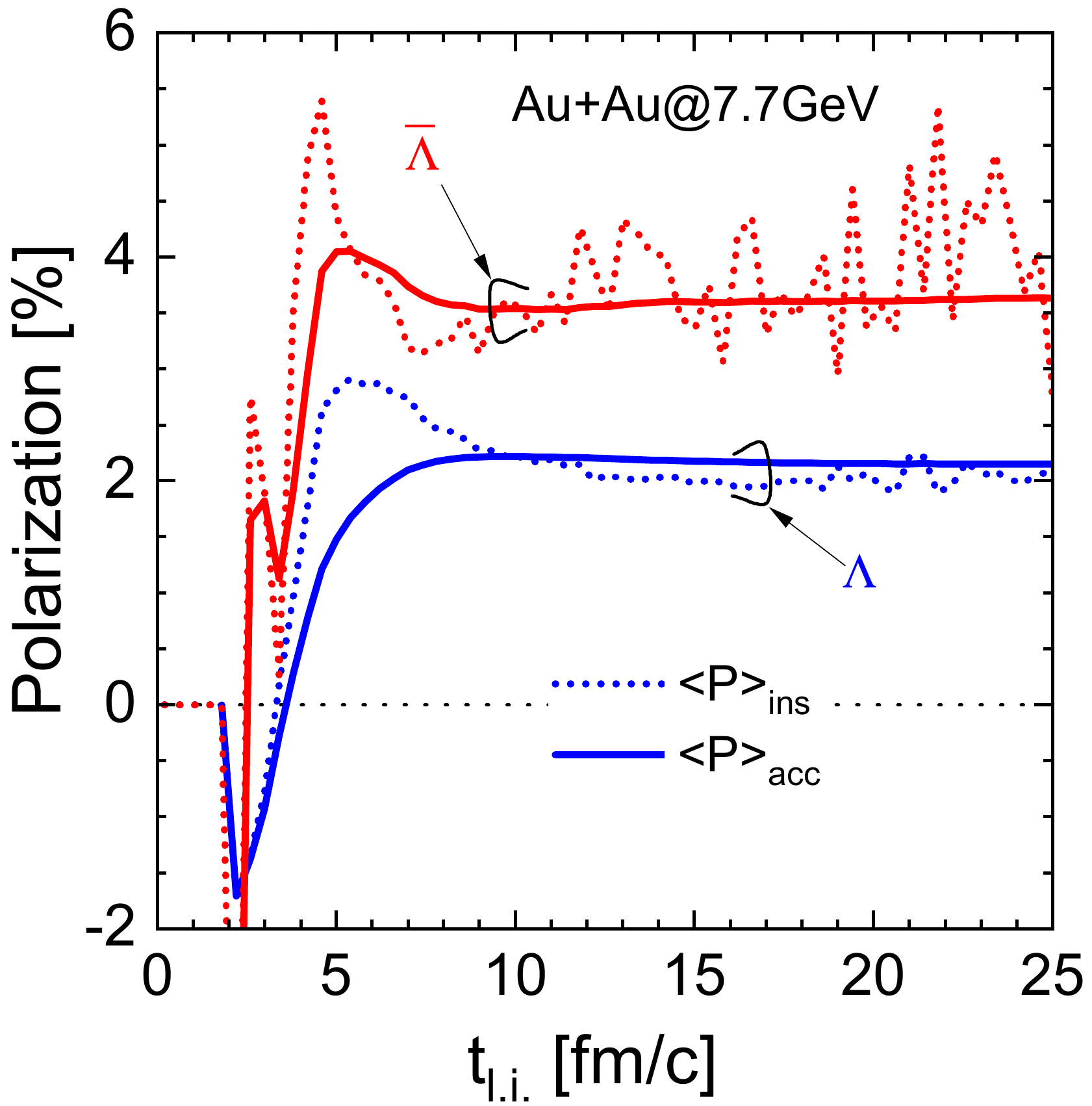}\\
\caption{Dotted lines show the instant polarization of $\Lambda$ and $\overline{\Lambda}$ hyperons released at the time moment $t_{\rm l.i.}$, see Eq.~(\ref{P-inst}). Solid lines stand for the polarization signal accumulated in the hyperons released until the time $t_{\rm l.i}$, see Eq.~(\ref{P-acc}). The calculation is done for Au+Au collision at energy $\sqrt{s_{NN}}=7.7$\,GeV and centrality range and the transverse momentum and rapidity cuts used by STAR experiment in~\cite{Adamczyk:2017}. At this collision energy, the touch time is 2.5\,fm/$c$ and the maximum overlap time is 5\,fm/$c$.
}
\label{fig:pol-prod}
\end{figure}

In Fig.~\ref{fig:pol:time} we show the distribution of the $\Lambda$ and $\overline{\Lambda}$ polarization (shown by the density histogram) for various TLI defined as
\begin{align}
w_{H}(t_{\rm l.i.},P)= \frac{1}{N_{\rm ev}}\frac{\rmd^2 N_{H}(t_{\rm l.i.},P)}{\rmd t_{\rm l.i.}\rmd P}\,,\,\,
H=\Lambda,\,\, \overline{\Lambda},
\label{dW-dist}
\end{align}
and normalized to the total hyperon multiplicity, cf. Eq.~(\ref{MH-def}),
\begin{align}
\int_0^{t_{\rm fin}}\rmd t_{\rm l.i.}\!\!\int_{-1}^{1}\rmd P \, w_{H}(t_{\rm l.i.},P)
=\int_0^{t_{\rm fin}}\frac{\rmd t_{\rm l.i.}}{N_{\rm ev}}\frac{\rmd N_H}{\rmd t_{\rm l.i.}}
 = M_{H,{\rm tot}}\,.
\end{align}
The integration of distribution (\ref{dW-dist}) over the polarization degree, $P$, gives the distribution shown in Fig.~\ref{fig:NTimeLi}a.
Calculations are done for Au+Au collisions at $\sqrt{s_{NN}}=7.7$\,GeV and events are selected according to the centrality range and the transverse momentum and rapidity cuts used in the STAR experiment~\cite{Adamczyk:2017}.
In the $\Lambda$ distribution we clearly recognize two `hot spots': the one at early times between the touch time 2.5\,fm/$c$ and the maximum overlap time 5\,fm/$c$, and the second one at later times $5\,{\rm fm}/c\lsim t\lsim 12$\,fm/$c$. In the $\overline{\Lambda}$ distribution there is only one hot spot at times $5\,{\rm fm}/c\lsim t\lsim 10$\,fm/$c$.
In Fig.~\ref{fig:pol:time} we see that for each TLI the polarization distributions are not symmetric which corresponds to a non-vanishing polarization of released hyperons. Nevertheless, the vast amount of $\Lambda$s is released with zero polarization during the times 10--20\,fm . These $\Lambda$s stem from rescattering processes occurring in the dilute matter with vanishing vorticity. In Fig.~\ref{fig:pol:time} they constitute an intensively dark line at zero polarization. For $\overline{\Lambda}$ the asymmetry of polarization distributions is larger.
The instant polarization carried by the (anti-)hyperons released at the time $t_{\rm l.i.}$, defined as
\begin{align}
\langle P_H(t_{\rm l.i.})\rangle_{\rm ins} =\Big[\frac{1}{N_{\rm ev}}\frac{\rmd N_H}{\rmd t_{\rm l.i}}\Big]^{-1} \!\!\int_{-1}^{1}\!\!\rmd P\,  P  \, w_{H}(t_{\rm l.i.},P),
\label{P-inst}
\end{align}
is shown in Fig.~\ref{fig:pol-prod} by dotted lines. At any time moment the instant polarization of the released $\overline{\Lambda}$s is higher than that of the released $\Lambda$s. Also, the polarization fluctuates more strongly for $\overline{\Lambda}$s. We see that the polarized hyperons are released even at times larger than the averaged freeze-out time, $t_{\Lambda(\overline{\Lambda})}^{\rm (f.o.)}$ and the time $t_{\Lambda(\overline{\Lambda})}^{\rm (80\%)}$. These hyperons stem from the Lorentz-delayed decays of resonances, their number is small but the degree of polarization is high.

The averaged spin polarization per hyperon accumulated in particles, $H=\Lambda$ and $\overline{\Lambda}$, released to the specific time moment $t_{\rm l.i.}$ can be calculated as
\begin{align}
\langle P_H(t_{\rm l.i.})\rangle_{\rm acc} &= \frac{1}{M_{H}(t_{\rm l.i.}) }
\int_{0}^{t_{\rm l.i.}}\!\!\rmd t'_{\rm l.i.} \int_{-1}^{1}\!\!\rmd P\, P\,
w_{H}(t'_{\rm l.i.},P) \,,
\label{P-acc}
\end{align}
where $M_{H}(t_{\rm l.i.})$ is given by Eq.~(\ref{MH-def}).
This quantity is depicted in Fig.~\ref{fig:pol-prod} by solid lines.
We obtain that for a collision at $\sqrt{s_{NN}}=7.7$\,GeV the signal of the spin polarisation is formed during the first 10\,fm/$c$ for both $\Lambda$ and $\overline{\Lambda}$ particles and stays approximately constant up to the full decomposition of the system. The variation of accumulated polarization is suppressed by the decrease of hyperon production rate at later times according to the equations
$
\frac{\rmd }{\rmd t_{\rm l.i.}}\langle P_H\rangle_{\rm acc}=
\frac{\dot{N}_H}{N_H} (\langle P_H\rangle_{\rm ins}-\langle P_H\rangle_{\rm acc})
$
following from definitions (\ref{P-inst}) and (\ref{P-acc}).
We see in Fig.~\ref{fig:pol:time} that the accumulated polarization of $\overline{\Lambda}$s is always larger than that of $\Lambda$s. Remarkably, that at a very early moment the instant and accumulated polarization both change the sign. This occurs in the time between the first touch of nuclei ($t\approx 2.2$\,fm/$c$) till they maximum overlap ($t \approx 4.9$\,fm/$c$).

\begin{figure}
\centering
\includegraphics[height=6cm]{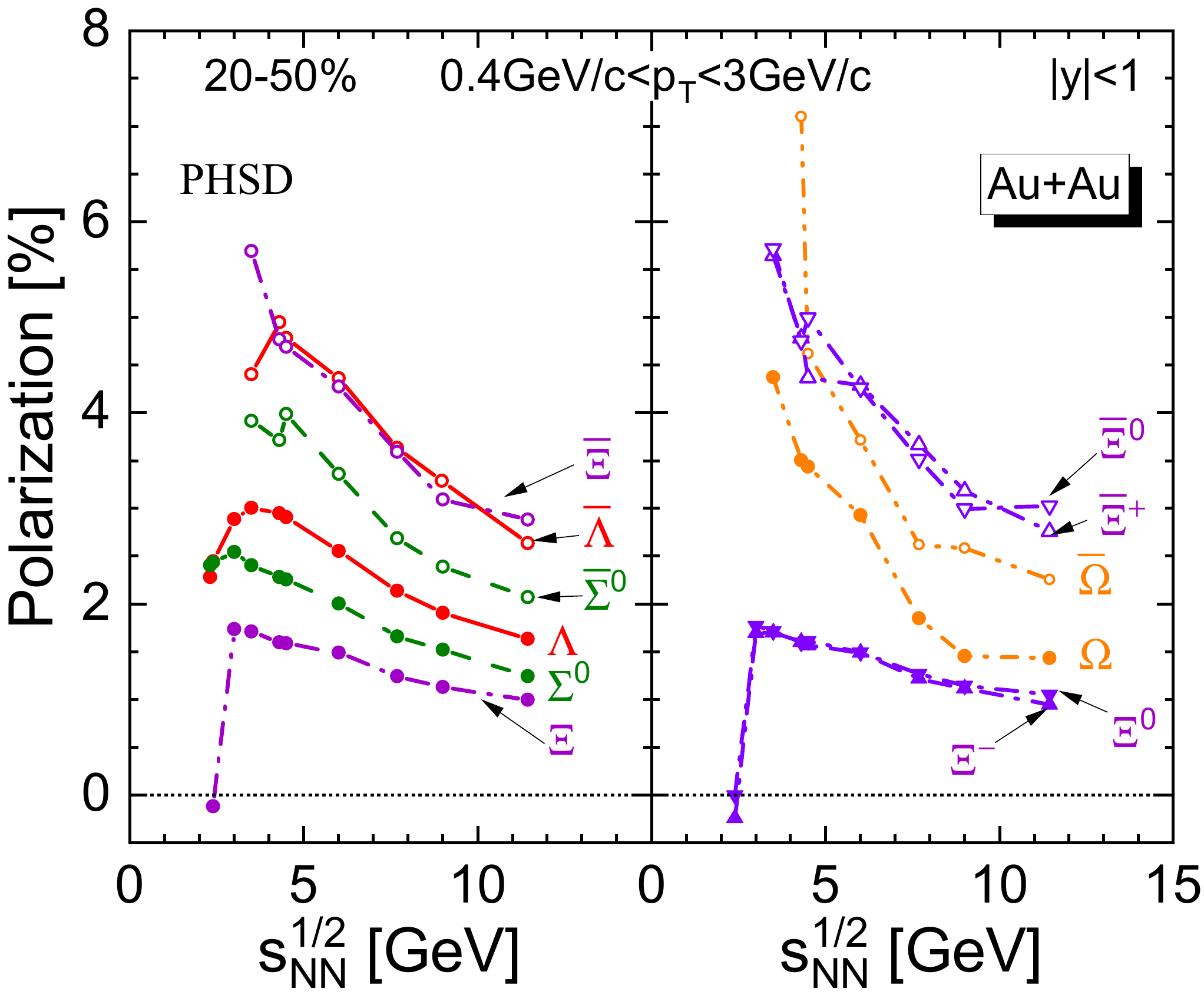}
\caption{Global polarization of different species of produced hyperons without weak and electromagnetic decays. The centrality class is $20-50\%$, the rapidity and transverse momentum regions are $|y|<1$ and $0.4{\rm GeV/}c<p_{\rm T}<3{\rm GeV/}c$, correspondingly.}
\label{fig:pol:hyp}
\end{figure}

The polarizations of various hyperon species produced in Au+Au collisions are shown in Fig.~\ref{fig:pol:hyp} as functions of the collision energy. For $\sqrt{s_{NN}}>4.5$\,GeV, the polarization of all hyperons decreases with an increase of the collision energy. For lower energies the $\Lambda$, $\Sigma^0$, and $\overline{\Sigma}^0$ the polarization starts falling with energy decrease since all polarization phenomena must vanish at the threshold energy $\sqrt{s_{NN}}=2m_N$ when nuclei do not move and the angular momentum of the system is zero. This drop of polarization for decreasing $\sqrt{s_{NN}}$ is not seen in Fig.~\ref{fig:pol:hyp} for  $\Xi(\overline{\Xi})$ and $\Omega(\overline{\Omega})$ since the number of these hyperons becomes very small and none of them fall within the acceptance range we impose.
We see also that the splitting between polarizations of particles and anti-particles exists not only for $\Lambda$s and $\overline{\Lambda}$s but also for other hyperon species. The following polarization hierarchy holds for collision energies between $\sqrt{s_{NN}}=3.5$\,GeV and 11.5\,GeV:
$P_{\overline{\Xi}}\simeq P_{\overline{\Lambda}} > P_{\overline{\Sigma}^0}>P_\Lambda >P_{\Sigma^0} >P_{\Xi}$.
The behavior of the $\Omega(\overline{\Omega})$ polarization with energy is quite different and does not follow the energy trend of other hyperons. While for all considered energies $P_{\overline{\Omega}}>P_{\Omega}$, both polarization rises with an energy decrease much more rapidly than that for other hyperons. So, at the collision energy 4.3\,GeV we have $P_{\Omega}=3.5\%$ and $P_{\overline{\Omega}}=7.1\%$ that is larger than, correspondingly, the polarization of $\Lambda$, $P_\Lambda=2.9\%$, and $\overline{\Lambda}$, $P_{\overline{\Lambda}}=4.9\%$. At the energy 11.5\,GeV we have oppositely $P_{\Omega(\overline{\Omega})}<P_{\Lambda(\overline{\Lambda})}$.

In comparison with the results obtained in Ref.~\cite{Wei:2019} within the AMPT model we obtain a similar polarization degree for $\Lambda$s , but by 30\% smaller polarization of $\Xi$s (1.8\% vs. 1.24\% in our calculations) and almost factor 2 smaller polarization for $\Omega$ hyperons (3.9\% vs 1.8\% in our case).

\section{Feed-down effects}\label{sec:feed-down}

Calculating the hyperon polarizations in the previous section we made an account only for strong decays of hyperons, which are built into the PHSD code. The final observed signal is formed, however, also by the weak decays
\begin{align}
\label{eq:feedown:weak}
\Xi \rightarrow \Lambda + \pi\,, \quad \Omega\to \Lambda + K^- \,,\quad \Omega \to \Xi +\pi\,,
\end{align}
and electromagnetic decays
\begin{align}\label{eq:feedown:ew}
\Sigma^0 \rightarrow \Lambda + \gamma.
\end{align}
The multiplicities of secondary $\Lambda$s and $\Xi^-$s are given by
\begin{align}
N_{\Lambda}^{(\rm sec)} &= N_{\Sigma^0} + B_{\Lambda\Xi}\: (N_{\Xi^0} + N_{\Xi^-}
+B_{\Xi\Omega}\, N_\Omega)
\nonumber\\
& + B_{\Lambda \Omega} N_\Omega \,,
\nonumber\\
N_{\Xi^-}^{\rm (sec)} &= B_{\Xi^-\Omega}\, N_\Omega\,,\quad N_{\Xi^0}^{\rm (sec)} = B_{\Xi^0\Omega}\, N_\Omega
\end{align}
Similar expressions hold for anti-hyperons.
The secondary component is mostly important for $\Lambda$ hyperons because of the significant contribution of $\Sigma^0$.
Contributions from $\Xi$ hyperons to $\Lambda$s and $\Omega$ hyperons to $\Xi$s are much smaller since $N_\Omega\ll N_\Xi\ll N_{\Sigma^0}$.
The same holds true also for corresponding anti-hyperons. The necessary branching ratios are $B_{\Lambda\Xi}=0.995$, $B_{\Xi\Omega} = B_{\Xi^0\Omega} +B_{\Xi^-\Omega}$, $B_{\Xi^0\Omega} = 0.236$, $B_{\Xi^-\Omega} = 0.086$, and $B_{\Lambda \Omega}=0.678$.

The primary $\Lambda$ polarization signal and the partial contributions of heavier hyperons to the measured $\Lambda$-polarization signal are
\begin{align}
\vec{S}_\Lambda^{*\rm (prim)} &= f_{\Lambda\Lambda}\, \vec{S}^*_{\Lambda},
\label{feed-down-Lprim}\\
\vec{S}_\Lambda^{*\rm (\Sigma^0)} &=  \kappa_{\Lambda}^{\Sigma^0}  \vec{S}^*_{\Sigma^0},
\,\, \kappa_{\Lambda}^{\Sigma^0} = f_{\Lambda\Sigma^0} C_{\Lambda\Sigma^0} ,
\label{feed-down-LSig}\\
\vec{S}_\Lambda^{*\rm (\Xi^{-,0})} &= \kappa_{\Lambda}^{\Xi^{-,0}}\vec{S}^*_{\Xi^{-0}},\,
\kappa_{\Lambda}^{\Xi^{-,0}} = f_{\Lambda \Xi^{-,0}}  C_{\Lambda\Xi^{-,0}} B_{\Lambda\Xi},
\label{feed-down-LXi}\\
\vec{S}_\Lambda^{*\rm (\Omega)} &=  \kappa_{\Lambda}^{\Omega}\,\vec{S}^*_\Omega , \,\,
\kappa_{\Lambda}^{\Omega} = f_{\Lambda\Omega}\big(C_{\Lambda\Xi^-}C_{\Xi^-\Omega}B_{\Xi^-\Omega}
\nonumber\\
&  + C_{\Lambda\Xi^0}C_{\Xi^0\Omega}B_{\Xi^0\Omega} + C_{\Lambda\Omega}B_{\Lambda\Omega}\big),
\label{feed-down-LOm}
\end{align}
and for the $\Xi^-$ and $\Xi^0$ signals, the contributions are
\begin{align}
\vec{S}_{\Xi^-}^{*\rm (prim)} &= f_{\Xi^-\Xi^-} \vec{S}^*_{\Xi^-},
\nonumber\\
\vec{S}_{\Xi^-}^{*(\Omega)} &= \kappa_{\Xi^-}^{\Omega} \vec{S}^*_\Omega \,, \quad
\kappa_{\Xi^-}^{\Omega} = f_{\Xi^-\Omega} C_{\Xi^-\Omega} B_{\Xi^-\Omega} ,
\nonumber\\
\vec{S}_{\Xi^0}^{*\rm (prim)} &= f_{\Xi^0\Xi^0}\, \vec{S}^*_{\Xi^0}\,,
\nonumber\\
\vec{S}_{\Xi^0}^{*(\Omega)} &= \kappa_{\Xi^0}^{\Omega} \vec{S}^*_\Omega,\quad
\kappa_{\Xi^0}^{\Omega} =  f_{\Xi^0\Omega} C_{\Xi^0\Omega} B_{\Xi^0\Omega} \,.
\label{feed-down-Xprim}
\end{align}
Here $f_{HH'}=N_{H'}/(N_H + N_H^{\rm (sec)})$ is the relative contribution of hyperon $H'$ to the final yield of hyperon $H$, and quantities $C_{HH'}$ are the spin transfer coefficients from the parent hyperon $H'$ to the daughter hyperon $H$. In Ref.~\cite{Becattini:2017} it was found
$C_{\Lambda \Sigma^{0}} =-1/3$. For weak decays in Eq.~(\ref{eq:feedown:weak}) the spin transfer coefficients are~\cite{Luk88,Bunce79}
\begin{align}
C_{\Lambda\Xi}={\textstyle\frac13}(1+2\gamma_\Xi)\,,\quad C_{H\Omega}=\textstyle{\frac15}(1+4\gamma_{H\Omega}) ,
\end{align}
where $\gamma_{\Xi(\Omega)}$ is the parameters of the weak baryon decay.
Using the values $\gamma_{\Xi^0}=0.87$ and $\gamma_{\Xi^-}=0.92$ from Ref.~\cite{PDG22} we obtain $C_{\Lambda \Xi^0}=0.914$, $C_{\Lambda \Xi^-} =0.943$. For the $\Omega$ baryons, the decay parameters $\gamma_{H\Omega}$, $H=\Lambda,\Xi$ are not determined experimentally.  In Ref.~\cite{Adam:2021} one argues that the choice $\gamma_{\Lambda \Omega}\simeq 1$ is preferable.  So, assuming that
$\gamma_{\Xi \Omega}\simeq 1$ also, we will use $C_{\Lambda\Omega}\simeq C_{\Xi^{-,0}   \Omega}\simeq 1$. Note, there are no feed-down contributions for the $\Omega$ hyperons. The relations similar to  Eqs.~(\ref{feed-down-Lprim}-\ref{feed-down-Xprim}) can be written for anti-hyperons with the same spin transfer coefficients, $C_{\overline{H}\overline{H'}}=C_{HH'}$ and branching ratios.

For $\sqrt{s_{NN}}=7.7$\,GeV the primary contribution for $\Lambda $ and $\overline{\Lambda}$ are
${|\vec{S}_\Lambda^{*\rm (prim)}|}/{|\vec{S}^*_{\Lambda}|} = 0.67$ and ${|\vec{S}_{\overline{\Lambda}}^{*\rm (prim)}|}/{|\vec{S}^*_{\overline{\Lambda}}|} = 0.51$, so the signal is substantially diluted by the secondary hyperons which carry smaller polarization. This reduction is stronger for  $\overline{\Lambda}$s than for $\Lambda$s. The contributions from other hyperons to the $\Lambda(\overline{\Lambda})$ spin polarization are controlled by the coefficients: $\kappa_{\Lambda}^{\Sigma^0} = -0.071$, $\kappa_{\Lambda}^{\Xi^0} =0.055$, $\kappa_{\Lambda}^{\Xi^-}=0.049$, and $\kappa_{\Lambda}^{\Omega} =1.6\times 10^{-3}$. These coefficients show that the $\Sigma^0$ contribution is negative and quite substantial, and is only in part compensated by $\Xi$ hyperons, since $P_{\Sigma^0} >P_\Xi$. The contribution of $\Omega$ hyperon can be safely neglected. For the anty-hyperons the coefficients for $\overline{\Sigma}^0$ increase by 30\%
$\kappa_{\overline{\Lambda}}^{\overline{\Sigma}^0} = -0.091$ and for $\overline{\Xi}$ the coefficients increase roughly by factor 2, $\kappa_{\overline{\Lambda}}^{\overline{\Xi}^0} = 0.10$ and $\kappa_{\overline{\Lambda}}^{\overline{\Xi}^-} =0.098$. For $\overline{\Omega}$, the coefficient remains very small, $\kappa_{\overline{\Lambda}}^{\overline{\Omega}}=9.1\times 10^{-3}$. For $\Xi(\overline{\Xi})$ hyperons the contributions from secondary processes are very small both for hyperons ${|\vec{S}_{\Xi^0}^{*\rm (prim)}|}/{|\vec{S}^*_{\Xi^0}|} =0.994$ and
${|\vec{S}_{\Xi^-}^{*\rm (prim)}|}/{|\vec{S}^*_{\Xi^-}|} =0.997$, and for anti-hyperons
${|\vec{S}_{\overline{\Xi}^0}^{*\rm (prim)}|}/{|\vec{S}^*_{\overline{\Xi}^0}|} = 0.980$ and
${|\vec{S}_{\overline{\Xi}^-}^{*\rm (prim)}|}/{|\vec{S}^*_{\overline{\Xi}^-}|} =0.992$, and, therefore, can be neglected.

\begin{figure}
\centering
\includegraphics[width=6cm]{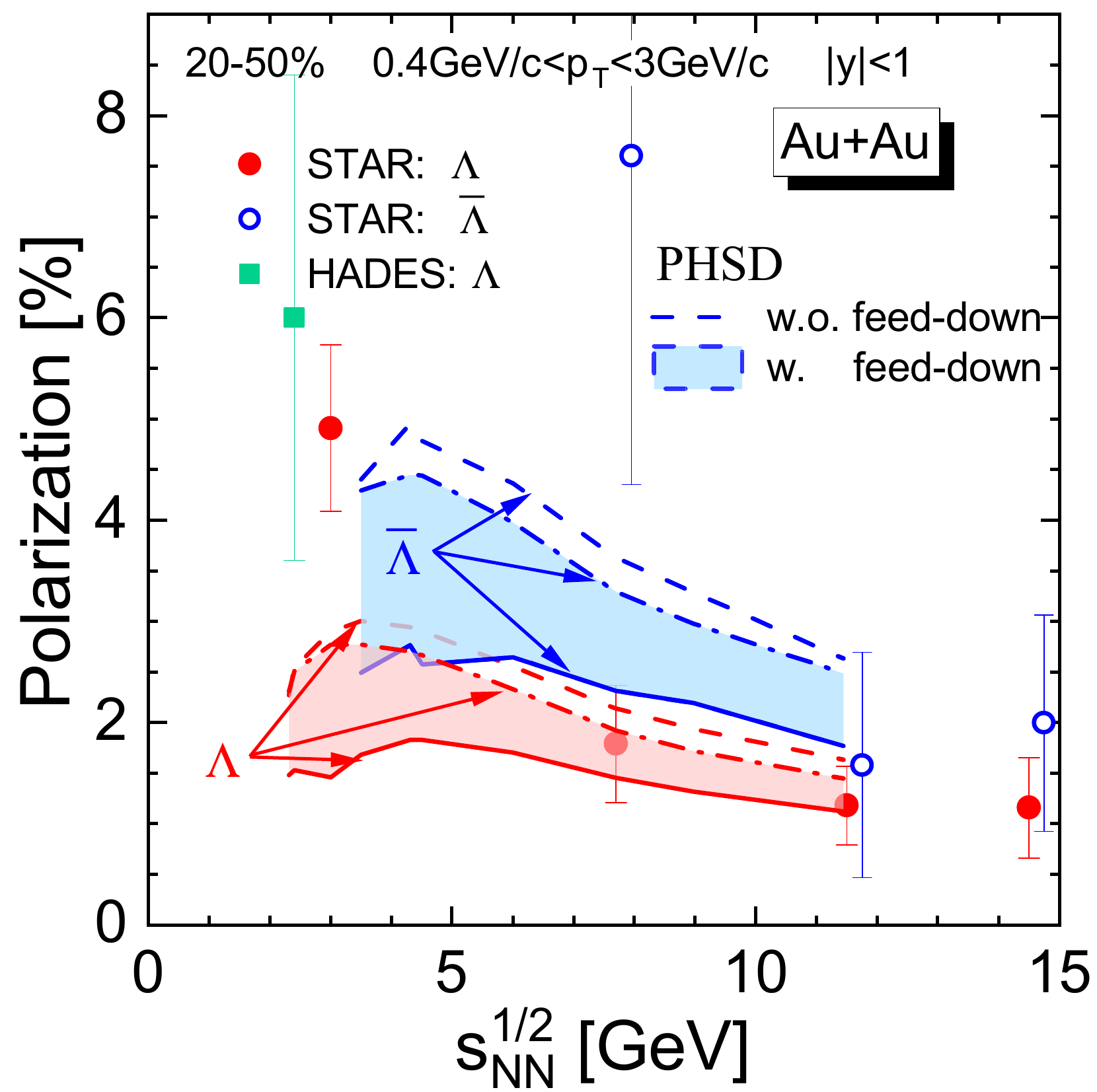}
\caption{Global polarizations of $\Lambda$ and $(\overline{\Lambda})$ hyperons in Au+Au collisions	at 20--50\% centrality, midrapidity $|y|<1$ and transverse momentum range $0.4\,{\rm GeV}/c < p_T < 3\,{\rm GeV}/c$ as functions of collision energy $\sqrt{s_{NN}}$. Dashed lines show hyperon polarization without feed-down procedure, solid lines refer to the results obtained taking into account decays (\ref{eq:feedown:weak}), (\ref{eq:feedown:ew}), and dot-dashed lines indicate the case of indistinguishability between $\Lambda$ and $\Sigma^0$ hyperons with the feeddown from $\Xi^0$, $\Xi^-$ decays, only. The STAR data~\cite{Adamczyk:2017,PhysRevC.104.L061901} and HADES~\cite{Yassine:2022} are also displayed. The HADES experiment acceptance differs from the depicted one in the figure and is limited by the 10--40\% centrality, the middle region of the central mass rapidity $-0.5<y_{\rm cm}<0.3$ and transverse momentum range $0.2\,{\rm GeV}/c<p_T<1.5\,{\rm GeV}/c$.}
\label{fig:pol}
\end{figure}

Finally, the total measured spin vector fo $\Lambda(\overline{\Lambda})$ is
\begin{align}
\vec{S}^{*\rm (meas)}_{\Lambda(\overline{\Lambda})} &= \vec{S}_{\Lambda(\overline{\Lambda})}^{*\rm (prim)} +\vec{S}_{\Lambda(\overline{\Lambda})}^{*\rm (\Sigma^0)}+\vec{S}_{\Lambda(\overline{\Lambda})}^{*\rm (\Xi)}\,.
\label{eq:becattini-feeddown-total-spin}
\end{align}
Polarization of primary Lambdas (without feed-down) $\vec{P^*}^{\rm (prim.)}_{\Lambda(\overline{\Lambda})} = 2\: \vec{S^*}_{\Lambda(\overline{\Lambda})}$
and measured (with electro-weak feed-down)  $\vec{P^*}^{\rm (meas)}_{\Lambda(\overline{\Lambda})} = 2\: \vec{S^*}^{\rm (meas)}_{\Lambda(\overline{\Lambda})}$
are shown in Fig.~\ref{fig:pol} by dashed and solid lines, respectively.
The significant feed-down effect is mainly caused by the contributions from electromagnetic decay of $\Sigma^0(\overline{\Sigma}^0)$ hyperon.
Reference~\cite{Becattini:2017} observed a relatively smaller feed-down effect, since the strong decays of hyperon resonances, which increase the number of strongly polarized $\Lambda(\overline{\Lambda})$s, were treated as a part of the feed-down. In our case, the strong decays are part of the PHSD code, and only weak and electromagnetic decays are named as feed-down processes.

Figure~\ref{fig:pol} demonstrates that the splitting between the $\Lambda$ and $\overline{\Lambda}$ polarization increases with a decrease of collision energy. Although we can describe the $\Lambda$ polarization at collision energies 7.7 and 11.5\, GeV and the $\overline{\Lambda}$ polarization at 11.5\,GeV both with and without feed-down effect taken into account, our calculations underpredict the $\overline{\Lambda}$ polarization at 7.7\,GeV and the $\Lambda$ polarization at energies $\sim 3$\,GeV. The lack of polarization could be probably filled by the inclusion of high-spin hyperon resonances, such as, e.g., $\Lambda^*(1520)$ --- the spin 5/2 resonances, which are not a part of the standard PHSD particle set. Also the other sources of the spin polarization such as the thermal shear and the analog of the spin-Hall effect~\cite{Becattini:2021,Liu:2021,Becattinit:2021:2,Liu:2021:2} are not yet implemented.

The depletion of the polarization signal for $\Lambda(\overline{\Lambda})$s depends on the multiplicity of $\Sigma^0(\overline{\Sigma}^0)$ hyperons. Despite of a good agreement of $\Lambda+\Sigma^0$ production with experimental data (see Fig.~\ref{fig:dNdY-Y0-LambdaXi}),
the multiplicities of $\Sigma$ hyperons and particularly of $\Sigma^0$ are not well constrained by the experimental data, separately.
To demonstrate the sensitivity of the feed-down effect to the $\Sigma^0$ multiplicity let us assume the extreme case, that all $\Sigma^0$s produced by the code should be actually $\Lambda$s, and replace all $\Sigma^0$ by $\Lambda$s in the calculations of the feed-down. The result of such a strong assumption is shown in Fig.~\ref{fig:pol} by dot-dashed lines. The filled area indicates the uncertainty in the fraction of $\Sigma^0$ within the sum $\Lambda +\Sigma^0$, which is fixed experimentally.

Contamination of the primary $\Lambda$ signal by decaying $\Sigma^0$s shadows the measured polarization effect. Therefore, it looks more attractive to consider the global polarization of $\Xi$ hyperons (and/or $\Lambda$ originated from their decay) which experimentally could be clearly identified and would carry direct information about the spin polarization of the fireball. Moreover, a part of $\Xi$ comes from $\Xi^*$ decays and carries by factor 5/3 stronger polarization than other $\Xi$s. Our calculations give
for $\sqrt{s_{NN}}=7.7$\,GeV
\begin{align}
P_{\Xi^0} = 1.22\%\,,\,\, P_{\Xi^-}= 1.27\%\,,\,\,
P_{\Xi^0+\Xi^-} = 1.24\%\,,
\nonumber\\
P_{\overline{\Xi}^0} = 3.68\%\,,\,\, P_{\overline{\Xi}^+}= 3.50\%\,,\,\,
P_{\overline{\Xi}^0+\overline{\Xi}^+} = 3.59\%,
\end{align}
and for $\sqrt{s_{NN}}=11.5$\,GeV
\begin{align}
P_{\Xi^0} &= 0.943\%\,,\,\, P_{\Xi^-}= 1.05\%\,,\,\,
P_{\Xi^0+\Xi^-} = 0.993\%\,,
\nonumber\\
P_{\overline{\Xi}^0} &= 2.76\%\,,\,\,\,\, P_{\overline{\Xi}^+}= 3.03\%\,,\,\,
P_{\overline{\Xi}^0+\overline{\Xi}^+} = 2.88\%.
\end{align}
On the experimental side, there are only estimations of the combined $\Xi^0+\Xi^-$ polarization at higher energies: $P_\Xi=1.3\pm 1\%$ at 27\,GeV~\cite{Alpatov:2020} and $P_\Xi=0.47\pm 0.33\%$ at 200\,GeV~\cite{Adam:2021}.

\section{Conclusions}\label{sec:conclusion}

We studied the hyperon production and formation of their global spin polarization in Au+Au collision at energies $2.3\,{\rm GeV}\le\sqrt{s_{NN}}\le 11.5\,{\rm GeV}$ within the PHSD transport code. We traced (anti-)hyperons, which escape the interaction area, to the last interaction point and determined thermodynamic characteristics and the vorticity field at this point. The local velocity of the fluid was calculated in the Landau frame after the proper separation of spectator nucleons, see Ref.~\cite{helicity}. The calculations show that the difference in polarizations between hyperons and anti-hyperons arises naturally and can be related to the difference in thermodynamic conditions and vorticity field of the medium wherefrom the particle is released, see also Ref.~\cite{TKV-particles-23}. The largest splitting is observed for $\Xi$ and $\overline{\Xi}$ hyperons.

We investigated the dependence of the polarization signal on the collision energy. Polarizations of all hyperon species decrease with an energy increase for energies $\sqrt{s_{NN}}>5$\,GeV, and the decrease is the strongest for $\Omega$ and $\overline{\Omega}$ hyperons. The maximum of $\Lambda$ and $\overline{\Lambda}$ polarization occurs at $\sqrt{s_{NN}}\sim 4$\,GeV.

The final measurable polarization signal of $\Lambda$ and $\overline{\Lambda}$ contains contributions from secondary weak and electromagnetic decays of hyperons. As a result, the primary signal gets diluted and the observed polarization degree decreases. The main destructive contribution is due to the electromagnetic $\Sigma^0\to \Lambda +\gamma$ decays in which the hyperon polarization changes sign. We analyzed the impact of the uncertainty in the relative number of produced $\Sigma^0$ to $\Lambda$ hyperons on the final global spin polarization of $\Lambda$ hyperons.
Despite this uncertainty our calculations of the $\Lambda$ polarization at energies $\sqrt{s}>7.7$\,GeV agree with experimental data.
The polarization of $\overline{\Lambda}$ at energies $\sqrt{s}<11.5$\,GeV and the polarization of $\Lambda$ at $\sqrt{s_{NN}}\sim 3$\,GeV remain underestimated.

We showed that the polarization signal for $\Xi$ and $\overline{\Xi}$ hyperons are much less influenced by the feed-down effects.  Therefore, the experimental study of the $\Xi(\overline{\Xi})$ polarization would be a better probe for the vortical dynamics in the course of nucleus collisions.

\section*{Acknowledgment}

We thank Yu.B. Ivanov and D.N. Voskresensky for enlightening discussions.
The research was supported by the ``Govorun'' computational cluster provided by the Laboratory of Information Technologies of the Joint Institute for Nuclear Research in Dubna, Moscow Region, Russia. The work was supported in part by the grant VEGA~1/0521/22.

%=====================================================


\begin{thebibliography}{10}

%%%%%%%%%%%%%%%%%%%%%%%%%%%%%%%%%%%%%%%%%%%
% EXPERIMENTAL POLARIZATION
%%%%%%%%%%%%%%%%%%%%%%%%%%%%%%%%%%%%%%%%%%%

\bibitem{Abelev:2007}
B.I.~Abelev, et al., STAR, Global polarization measurement in Au+Au collisions, Phys. Rev. C \textbf{76} (2007) 024915, Erratum: Phys. Rev. C \textbf{95} (2017) 039906.

\bibitem{Adamczyk:2017}
L.~Adamczyk, et al., STAR, Global $\Lambda$ hyperon polarization in nuclear collisions: evidence for the most vortical fluid, Nature \textbf{ 548} (2017) 62.

\bibitem{PhysRevC.104.L061901}
M.S.~Abdallah, et al., STAR, Global $\Lambda$-hyperon polarization in Au+Au collisions at $\sqrt{{s}_{\mathrm{NN}}}=3$\,GeV, Phys. Rev. C \textbf{104} (2021) L061901.

\bibitem{Adam:2021}
J.~Adam, et al., STAR, Global polarization of $\Xi$ and $\Omega$ hyperons in Au+Au Collisions at $\sqrt{{s}_{NN}}=200$\,GeV, Phys. Rev. Lett. \textbf{126} (2021) 162301.

\bibitem{Yassine:2022}
R.A.~Yassine, et al., HADES, Measurement of global polarization of $\Lambda$ hyperons in few-GeV heavy-ion collisions, Phys. Lett. B \textbf{835} (2022) 137506.

\bibitem{DeGrand-81}
T.~De~Grand, H.~Miettinen, Quark dynamics of polarization in inclusive hadron production Phys. Rev. D \textbf{23}, 1227 (1981).

\bibitem{Becattini:2013}
F.~Becattini, V.~Chandra, L.~Del Zanna, E.~Grossi, Relativistic distribution function for particles with spin at local thermodynamical equilibrium, Annals Phys. \textbf{338}, 32 (2013).

\bibitem{Becattini:2015ska}
F.~Becattini, G.~Inghirami, V.~Rolando, A.~Beraudo, L.~Del Zanna, A.~De Pace, M.~Nardi, G.~Pagliara, V.~Chandra, A study of vorticity formation in high energy nuclear collisions, Eur. Phys. J. C \textbf{75}, 406 (2015).

\bibitem{PhysRevC77.024906}
F. Becattini, F. Piccinini, J. Rizzo, Angular momentum conservation in heavy ion collisions at very high energy, Phys. Rev. C \textbf{ 77}, 024906 (2008).

\bibitem{Barnett}
S.J. Barnett, Magnetization by Rotation, Phys. Rev. \textbf\textbf{6}, 239 (1915).

\bibitem{Karpenko:2017}
Iu.~Karpenko, F.~Becattini, Study of $\Lambda$ polarization in relativistic nuclear collisions at $\sqrt{s_{NN}}=7.7$--200 GeV, Eur. Phys. J. C \textbf{77} (2017) 213.

%\bibitem{Becattini-Inghirami-2015}
%F.~Becattini, G.~Inghirami, V.~Rolando, A.~Beraudo, L.~Del Zanna, A.~De Pace, M.~Nardi, G.~Pagliara, V.~Chandra,
%A study of vorticity formation in high energy nuclear collisions, Eur. Phys. J. C \textbf{75}, 406 (2015).

\bibitem{Xie:2017}
Y.~Xie, D.~Wang, L.P.~Csernai, Global $\Lambda$ polarization in high energy collisions, Phys. Rev. C \textbf{95} (2017) 031901(R).


\bibitem{Ivanov-Soldatov-2017}
Yu.B.~Ivanov, A.A.~Soldatov, Vorticity in heavy-ion collisions at the JINR Nuclotron-based Ion Collider fAcility,
Phys. Rev. C \textbf{95}, 054915 (2017)

\bibitem{Ivanov-Toneev-Soldatov-2019}
Yu.B.~Ivanov, V.D.~Toneev, A.A.~Soldatov, Estimates of hyperon polarization in heavy-ion collisions at collision energies
$\sqrt{s_{NN}}= 4$--40\,GeV, Phys. Rev. C \textbf{100}, 014908 (2019).

\bibitem{Ivanov:2020}
Yu.B.~Ivanov, Global polarization in heavy-ion collisions based on the axial vortical effect, Phys. Rev. C \textbf{102} (2020) 044904.

\bibitem{Ivanov:2021}
Yu.B.~Ivanov, Global $\Lambda$ polarization in moderately relativistic nuclear collisions, Phys. Rev. C \textbf{103} (2021) L031903.

\bibitem{Liu:2021}
S.Y.F.~Liu, Y.~Yin, Spin polarization induced by the hydrodynamic gradients, JHEP \textbf{07} (2021) 188.

\bibitem{Becattinit:2021:2}
F.~Becattini, M.~Buzzegoli, A.~Palermo, G.~Inghirami, I.~Karpenko, Local polarization and isothermal local equilibrium in relativistic heavy-ion collisions, Phys. Rev. Lett. \textbf{127} (2021) 272302.

\bibitem{Becattini:2021}
F.~Becattini, M.~Buzzegoli, A.~Palermo, Spin-thermal shear coupling in a relativistic fluid, Phys. Lett. B \textbf{ 820} (2021) 136519.

\bibitem{Liu:2021:2}
S.Y.F.~Liu, Y.~Yin, Spin Hall effect in heavy-ion collisions, Phys. Rev. D \textbf{104} (2021) 054043.

\bibitem{Ivanov-zhetp}
Yu.B. Ivanov, A.A. Soldatov, On the ambiguity of the definition of shear and spin-hall
contributions to the $\Lambda$ polarization in heavy-ion collisions, JETP Lett. \textbf{116}, 133 (2022).


\bibitem{Sun:2017}
Y.~Sun, C.M.~Ko, $\Lambda$ hyperon polarization in relativistic heavy ion collisions from a chiral kinetic approach, Phys. Rev. C \textbf{96} (2017) 024906

%%%%%%%%%%%%%%%%%%%%%%%%%%%%%%%%%%%%%%%%%%%
% POLARIZATION CALCULATION
%%%%%%%%%%%%%%%%%%%%%%%%%%%%%%%%%%%%%%%%%%%

\bibitem{Li:2017}
H.~Li, L.-G.~Pang, Q.~Wang, X.-L.~Xia, Global $\Lambda$ polarization in heavy-ion collisions from a transport model, Phys. Rev. C \textbf{96} (2017) 054908.

\bibitem{Wei:2019}
D.-X.~Wei, W.-T.~Deng, X.-G.~Huang, Thermal vorticity and spin polarization in heavy-ion collisions, Phys. Rev. C \textbf{99} (2019) 014905.

\bibitem{Li-Xia-Huang-Huang-22}
Hui~Li, Xiao-Liang~Xia, Xu-Guang~Huang, Huan Zhong Huang, Global spin polarization of multistrange hyperons and feed-down effect in heavy-ion collisions, Phys. Lett. B \textbf{827}, 136971 (2022).

\bibitem{Han-Xu-18}
Zhang-Zhu~Han, Jun~Xu, Investigating different $\Lambda$ and $\overline{\Lambda}$ polarizations in relativistic heavy-ion collisions,
Phys. Lett. B \textbf{786}, 255 (2018).

\bibitem{Xu-Lin-Huang-Huang-22}
Kun~Xu, Fan~Lin, Anping~Huang, Mei~Huang, $\Lambda/\overline{\Lambda}$  polarization and splitting induced by rotation and magnetic field, Phys. Rev. D \textbf{106}, L071502 (2022)

\bibitem{Deng-Huang-Ma-22}
Xian-Gai Deng, Xu-Guang Huang, Yu-Gang Ma, Lambda polarization in $^{108}$Ag+$^{108}$Ag and $^{197}$Au+$^{197}$Au collisions around a few GeV,
Phys. Lett. B \textbf{835}, 137560 (2022).

\bibitem{VITIUK2020135298}
O.~Vitiuk, L.V.~Bravina, E.E.~Zabrodin, Is different $\Lambda$ and $\overline{\Lambda}$ polarization caused by different spatio-temporal freeze-out picture?, Phys. Lett. B \textbf{ 803}, 135298 (2020).

\bibitem{BGST-Hseparation}
M.I.~Baznat, K.K.~Gudima, A.S.~Sorin, O.V.~Teryaev, Helicity separation in heavy-ion collisions, Phys. Rev. C \textbf{  88}, 061901(R) (2013).

\bibitem{BGST-Vsheet}
M.I.~Baznat, K.K.~Gudima, A.S.~Sorin, O.V.~Teryaev, Femto-vortex sheets and hyperon polarization in heavy-ion collisions, Phys. Rev. C \textbf{93}, 031902 (2016).

\bibitem{PhysRevC.97.064902}
E.E.~Kolomeitsev, V.D.~Toneev, V.~Voronyuk, Vorticity and hyperon polarization at energies available at JINR Nuclotron-based Ion Collider fAcility, Phys. Rev. C \textbf{97} (2018) 064902.
	
\bibitem{Ivanov-rings}
Yu. B. Ivanov, A. A. Soldatov, Vortex rings in fragmentation regions in heavy-ion collisions at $\sqrt{{s}_{NN}}=39$ GeV, Phys. Rev. C \textbf{97}, 044915 (2018).

\bibitem{helicity}
N.S. Tsegelnik, E.E. Kolomeitsev, V. Voronyuk, Helicity and vorticity in heavy-ion collisions at energies available at the JINR Nuclotron-based Ion Collider facility, Phys. Rev. C \textbf{107}, 034906 (2023).

\bibitem{Voskre-mag}
D.N.~Voskresensky, N.Yu.~Anisimov, Properties of a pion condensate in a magnetic field, Zh. Eksp. Teor. Fiz. \textbf{78}, 2845 (1980)
[Sov. Phys. JETP \textbf{51}, 13 (1980)].

\bibitem{SIT-2009}
V.~Skokov, A.Y.~Illarionov, V.D.~Toneev, Estimate of the magnetic field strength in heavy-ion collisions,
Int. J. Mod. Phys. A \textbf{24}, 5925  (2009).

\bibitem{Becattini:2017}
F.~Becattini, I.~Karpenko, M.A.~Lisa, I.~Upsal, S.A.~Voloshin, Global hyperon polarization at local thermodynamic equilibrium with vorticity, magnetic field, and feed-down, Phys. Rev. C \textbf{95} (2017) 054902.

%\cite{Buzzegoli:2022qrr}
\bibitem{Buzzegoli-mag-2022}
M.~Buzzegoli,
Spin polarization induced by magnetic field and the relativistic Barnett effect,
arXiv:2211.04549.
%%%%%%%%%%%%%%%%%%%%%%%%%%%%%%%%%%%%%%%%%%%%%

\bibitem{Rogachevsky-ST-2010}
O.V.~Rogachevsky, A.S.~Sorin, O.V.~Teryaev, Chiral vortaic effect and neutron asymmetries in heavy-ion collisions, Phys. Rev. C \textbf{82}, 054910 (2010).

\bibitem{Gao:2012}
J.-H.~Gao, Z.-T.~Liang, S.~Pu, Q.~Wang, X.-N.~Wang, Chiral anomaly and local polarization effect from the quantum kinetic approach, Phys. Rev. Lett. \textbf{ 109} (2012) 232301.

\bibitem{Baznat:2018}
M.~Baznat, K.~Gudima, A.~Sorin, O.~Teryaev, Hyperon polarization in heavy-ion collisions and holographic gravitational anomaly, Phys. Rev. C \textbf{97} (2018) 041902(R).

\bibitem{Csernai:2019}
L.~Csernai, J.~Kapusta, T.~Welle, $\Lambda$ and $\overline{\Lambda}$ spin interaction with meson fields generated by the baryon current in high energy nuclear collisions, Phys. Rev. C \textbf{99} (2019) 021901.

\bibitem{Xie-Chen-Csernai-2021}
Yilong Xie, Gang Chen, L.P.~Csernai, A study of $\Lambda$ and $\overline{\Lambda}$ polarization splitting by meson field in PICR
hydrodynamic model, Eur. Phys. J. C  \textbf{81}, 12 (2021).

\bibitem{Ivanov:2022}
Yu.B.~Ivanov, A.A.~Soldatov, Global $\Lambda$ polarization in heavy-ion collisions at energies 2.4--7.7\,GeV: Effect of meson-field interaction, Phys. Rev. C \textbf{105} (2022) 034915.

\bibitem{Ladygin:2010}
V.P.~Ladygin, A.P.~Jerusalimov, N.B.~Ladygina, Polarization of $\Lambda^0$ hyperons in nucleus-nucleus collisions at high energies, Phys. Part. Nuclei Lett. \textbf{7} (2010) 349.

\bibitem{Ayala-PLB810}
A.~Ayala, et al., Core meets corona: A two-component source to explain $\Lambda$ and $\overline{\Lambda}$ global polarization in semi-central heavy-ion collisions, Phys. Lett. B \textbf{810} (2020) 135818.

\bibitem{Ayala-Particles23}
A.~Ayala, I.~Dominguez, I.~Maldonado, M.E.~Tejeda-Yeomans,
An improved core-corona model for $\Lambda$ and $\overline{\Lambda}$ polarization in relativistic heavy-ion collisions,
Particles \textbf{2023}, 405-415 (2023).
%https://doi.org/10.3390/particles6010022

\bibitem{Deng-Huang-Ma-Zhang-20}
Xian-Gai~Deng, Xu-Guang~Huang, Yu-Gang~Ma, Song~Zhang, Vorticity in low-energy heavy-ion collisions,
Phys. Rev. C \textbf{101}, 064908 (2020)

\bibitem{TKV-particles-23}
N.S. Tsegelnik, E.E. Kolomeitsev, V. Voronyuk, $\Lambda$ and $\overline{\Lambda}$ freeze-out distributions and global polarizations in Au+Au collisions, Particles \textbf{2023}, 373 (2023).
%https://doi.org/10.3390/particles6010019

%%%%%%%%%%%%%%%%%%%%%%%%%%%%%%%%%%%%%%%%%%%
%%%%%%%%%%%%%%%%%%%%%%%%%%%%%%%%%%%%%%%%%%%
%%%%%%%%%%%%%%%%%%%%%%%%%%%%%%%%%%%%%%%%%%%

%\bibitem{PRL120:062301}
%	L. Adamczyk et al. (STAR Collaboration), Beam-Energy Dependence of Directed Flow of $\Lambda$, $\overline{\Lambda}$, ${K}^{\pm}$, ${K}_{s}^{0}$, and $\phi$ in Au+Au Collisions, Phys. Rev. Lett. \textbf{ 120}, 062301 (2018).

%\bibitem{PhysRevC.93.014907}
%	L. Adamczyk et al. (STAR Collaboration), Centrality dependence of identified particle elliptic flow in relativistic heavy ion collisions at $\sqrt{{s}_{NN}}=7.7--62.4$ GeV,Phys. Rev. C \textbf{ 93}, 014907 (2016).

%\bibitem{Munzinger-Stachel:1998}
%	P. Braun-Munzinger and J. Stachel, Dynamics of ultra-relativistic nuclear collisions with heavy beams: An experimental overview, Nucl. Phys. A {\bf638}, 3 (1998).
%
%\bibitem{Pinkenburg:1999}
%	C. Pinkenburg et al. (E895 Collaboration), Elliptic Flow: Transition from Out-of-Plane to In-Plane Emission in $\mathrm{Au}+\mathrm{Au}$ Collisions, Phys. Rev. Lett. {\bf83}, 1295 (1999).
%
%\bibitem{Alt:2003}
%	C. Alt et al. (NA49 Collaboration), Directed and elliptic flow of charged pions and protons in $\mathrm{Pb}+\mathrm{Pb}$ collisions at $40A$ and $158A\mathrm{GeV}$, Phys. Rev. C {\bf68}, 034903 (2003).
%	
%\bibitem{Adamova:2005}
%	D. Adamova et al. (CERES Collaboration), New results from CERES, Nucl. Phys. A {\bf698}, 253 (2002).
%
%\bibitem{ANDRONIC2005173}
%	A. Andronic et al. (FOPI Collaboration), Excitation function of elliptic flow in Au+Au collisions and the nuclear matter equation of state, Physics Letters B {\bf612}, 173180 (2005).
%
%\bibitem{Adamczyk:2012}
%	L. Adamczyk et al. (STAR Collaboration), Inclusive charged hadron elliptic flow in Au $+$ Au collisions at $\sqrt{{s}_{NN}}=7.7-39\,$GeV, Phys. Rev. C {\bf86}, 054908 (2012).

%%%%%%%%%%%%%%%%%%%%%%%%%%%%%%%%%%%%%%%%%%%
% NICA
%%%%%%%%%%%%%%%%%%%%%%%%%%%%%%%%%%%%%%%%%%%

%\bibitem{Kekelidze:2017}
%	V. D. Kekelidze, V. A. Matveev, I. N. Meshkov, A. S. Sorin, G. V. Trubnikov, Project Nuclotron-based Ion Collider fAcility at JINR, Phys. Part. Nucl. \textbf{ 48}, 727 (2017).


%%%%%%%%%%%%%%%%%%%%%%%%%%%%%%%%%%%%%%%%%%%
% PHSD
%%%%%%%%%%%%%%%%%%%%%%%%%%%%%%%%%%%%%%%%%%%
\bibitem{PHSD-I}
W. Cassing, E. L. Bratkovskaya, Parton-hadron-string dynamics:
An off-shell transport approach for relativistic energies,
Nucl. Phys. A 831, 215 (2009).

\bibitem{PHSD-II}
E.L.~Bratkovskaya, W.~Cassing, V.P.~Konchakovski, O. Linnyk, Parton-hadron-string dynamics at relativistic collider energies,
Nucl. Phys. A \textbf{856}, 162 (2011).

\bibitem{PHSD-III}
O. Linnyk, E. Bratkovskaya, W. Cassing, Effective QCD and transport description of dilepton and photon production in
heavy-ion collisions and elementary processes, Prog. Part. Nucl. Phys. \textbf{87}, 50 (2016).

\bibitem{Cassing:2015owa}
W.~Cassing, A.~Palmese, P.~Moreau, E.~L. Bratkovskaya, Chiral symmetry restoration versus deconfinement in heavy-ion collisions at high baryon density, Phys. Rev. C \textbf{93}, 014902 (2016).

\bibitem{PHSD-IV}
A. Palmese, W. Cassing, E. Seifert, T. Steinert, P. Moreau, E.L. Bratkovskaya,
Chiral symmetry restoration in heavy-ion collisions at intermediate energies, Phys. Rev. C \textbf{94}, 044912 (2016).

%%%%%%%%%%%%%%%%%%%%%%%%%%%%%%%%%%%%%%%%%%%
% EXPERIMENT. YIELDS
%%%%%%%%%%%%%%%%%%%%%%%%%%%%%%%%%%%%%%%%%%%
\bibitem{PhysRevB.386.034909}
S.~Ahmad, et al., (E891 Collaboration), $\Lambda$ production by 11.6\,GeV/$c$ Au beam on Au target, Phys. Lett. B \textbf{  382}, 35 (1996), Erratum: Phys. Lett. B \textbf{  386}, 496 (1996).

\bibitem{PhysRevLett.87.242301}
B.B.~Back, et al., (E917 Collaboration), Antilambda Production in Au+Au Collisions at 11.7\,$A$\,GeV/$c$, Phys. Rev. Lett. \textbf{  87}, 242301 (2001).

\bibitem{PhysRevLett.88.062301}
S. Albergo, et al., (E896 Collaboration), $\Lambda$ Spectra in 11.6$A$\,GeV/$c$ Au-Au Collisions, Phys. Rev. Lett. \textbf{  88}, 062301 (2002).

\bibitem{PhysRevC.78.034918}
C. Alt, et al., (NA49 Collaboration), Energy dependence of $\Lambda$ and $\Xi$ production in central Pb+Pb collisions at 20$A$, 30$A$, 40$A$, 80$A$, and 158$A$\,GeV measured at the CERN Super Proton Synchrotron, Phys. Rev. C \textbf{78}, 034918 (2008).

\bibitem{PhysRevC.102.034909}
J. Adam, et al., (STAR Collaboration), Strange hadron production in Au+Au collisions at $\sqrt{s_{NN}}=$ 7.7, 11.5, 19.6, 27, and 39\,GeV, Phys. Rev. C \textbf{102}, 034909 (2020).

\bibitem{NA49-Alt-Omega}
C. Alt, et al., (NA49 Collaboration), $\Omega^-$ and $\overline{\Omega}^+$ production in central Pb+Pb collisions at 40 and 158$A$\,GeV
Phys. Rev. Lett. \textbf{94}, 192301 (2005).
%%%%%%%%%%%%%%%%%%%%%%%%%%%%%%%%%%%%%%%%%%%
% STATISTICAL APPROACH TO POLARIZATION
%%%%%%%%%%%%%%%%%%%%%%%%%%%%%%%%%%%%%%%%%%%
%\bibitem{Becattini:2013fla}
%F.~Becattini, V.~Chandra, L.~Del Zanna, and E.~Grossi, Relativistic distribution function for particles with spin at local thermodynamical equilibrium, Annals Phys. \textbf{ 338}, 32 (2013)


%%%%%%%%%%%%%%%%%%%%%%%%%%%%%%%%%%%%%%%%%%%
% OTHER
%%%%%%%%%%%%%%%%%%%%%%%%%%%%%%%%%%%%%%%%%%%
\bibitem{SDM09}
L. M. Satarov, M. N. Dmitriev, I. N. Mishustin, Equation of state of hadron resonance gas and the phase diagram of strongly interacting matter, Phys. Atom. Nucl. \textbf{72}, 1390 (2009).


\bibitem{PDG22}
R.L. Workman, et al., (Particle Data Group), Review of particle physics, Prog. Theor. Exp. Phys. \textbf{2022}, 083C01 (2022)

\bibitem{Luk88}
K.B.~Luk, A.~Beretvas, L.~Deck, et al., New measurements of properties of the $\Omega^-$ hyperon,  Phys. Rev. D \textbf{38}, 19 (1988)

\bibitem{Bunce79}
G.~Bunce O.E.~Overseth, P.T.~Cox, et al., The first measurement of the $\Xi^0$ magnetic moment, Phys. Lett. B \textbf{86}, 386 (1979)

\bibitem{Alpatov:2020}
E.~Alpatov (for the STAR collaboration), Global hyperon polarization in Au+Au collisions at $\sqrt{{s}_{NN}}=27\,{\rm{GeV}}$ in the STAR experiment, J. Phys.: Conf. Ser. \textbf{1690}, 012120 (2020).

\end{thebibliography}
\end{document}